\documentclass{aa}  
\usepackage{natbib}
\usepackage{graphicx}
\usepackage{txfonts}
\usepackage{hyperref}
\newcommand{\fso}{\ensuremath{f_0^*}\xspace}
\newcommand{\denv}{\ensuremath{d_\mathrm{env}}\xspace}
\newcommand{\modif}[1]{#1}
\newcommand{\modiff}[1]{#1}

\begin{document} 
  \title{VLTI/PIONIER reveals the close environment of the evolved system \object{HD101584}\thanks{Based on observations collected at the European Organisation for Astronomical Research in the Southern Hemisphere under ESO programme 099.D-0088}}
  \author{J. Kluska\inst{1}
        \and
          H. Olofsson\inst{2}
        \and
         H. Van Winckel\inst{1}
        \and 
           T. Khouri\inst{2}
          \and
            M. Wittkowski\inst{3} 
             \and
            W. J. de Wit\inst{4}
            \and 
           E. M. L. Humphreys\inst{3}
            \and 
            M. Lindqvist\inst{2}
            \and 
            M. Maercker\inst{2}
           \and 
           S. Ramstedt\inst{5} 
            \and 
            D. Tafoya\inst{2,6}
            \and
           W. H. T. Vlemmings\inst{2}
          }

  \institute{Instituut voor Sterrenkunde (IvS), KU Leuven, Celestijnenlaan 200D, 3001, Leuven, Belgium\\ \email{jacques.kluska@kuleuven.be}
         \and
    Department of Space, Earth and Environment, Chalmers University of Technology, Onsala Space Observatory, 43992, Onsala, Sweden
    \and
    ESO, Karl-Schwarzschild-Str. 2, 85748, Garching bei München, Germany
            \and
            ESO, Alonso de Cordova 3107, Vitacura, Santiago, Chile
            \and 
            Department of Physics and Astronomy, Uppsala University, Box 516, 75120, Uppsala, Sweden
            \and
            National Astronomical Observatory of Japan, 2-21-1 Osawa, Mitaka, Tokyo 181-8588, Japan
             }

  \date{Received XXX; accepted XXX}
  
\authorrunning{Kluska et al.}
\titlerunning{HD101584}

  \abstract
   {
   The observed orbital characteristics of post-asymptotic giant branch (post-AGB) and post-red giant branch (post-RGB) binaries are not understood. 
   We suspect that the missing ingredients to explain them probably lie in the continuous interaction of the central binary with its circumstellar environment.}
   {We aim at studying the circumbinary material in these complex systems by investigating the connection between the innermost and large-scale structures.}
   {We perform high-angular resolution observations in the near-infrared continuum of HD\,101584, that has a complex structure as seen at millimeter wavelengths with a disk-like morphology and a bipolar outflow \modif{due to an episode of a strong binary interaction}. To account for the complexity of the target we first perform an image reconstruction and use this result to fit a geometrical model to extract the morphological and thermal features of the environment.}
   {The image reveals an unexpected double ring structure. We interpret the inner ring to be produced by emission from dust located in the plane of the \modif{disk} and the outer ring to be produced by emission from dust that is located 1.6\,[D/1kpc]\,au above the \modif{disk} plane.
   The inner ring diameter (3.94\,[D/1kpc]\,au) and temperature (T=1540$\pm$10\,K) are compatible with the dust sublimation front of the disk. The origin of the out-of-plane ring (with a diameter of 7.39\,[D/1kpc]\,au and a temperature of 1014$\pm10$\,K) \modif{could be} due to episodic ejection or a dust condensation front in the outflow.}
   {The observed outer ring is possibly linked with the blue-shifted side of the large scale outflow seen by the Atacama Large Millimeter/submillimeter Array (ALMA) and is tracing its launching location to the \modif{central star}. Such observations give morphological constraints on the ejection mechanism. Additional observations are needed to constrain the origin of the out-of-plane structure.}

  \keywords{ Stars: AGB and post-AGB, Stars: binaries, Stars: circumstellar matter, Stars: winds, outflows, Techniques: high angular resolution, Techniques: interferometric}

\maketitle


\section{Introduction}

\modif{Binarity is frequent in stars and can strongly impact their evolution} \citep{Duchene2013,Sana2014}.
When the separation on the main sequence is of the order of an astronomical unit, binary interaction will severely impact the stellar evolution when the radius of the star increases giving rise to various types of objects and phenomena (such as barium stars, asymmetric planetary nebulae, and supernovae type Ia).
In \modiff{this paper} we focus on low- and intermediate mass stars with main sequence masses between about 0.8 and 8\,M$_\odot$. 
The stellar radius increases dramatically during the red giant and asymptotic giant phases, and, depending on the binary separation, this will lead to strong interaction via tides, wind-Roche lobe overflow \citep[e.g.][]{Abate2013} or full Roche lobe overflow \citep[e.g.][]{Ivanova2013}. 
Stellar evolution is, hence, different than that for single stars.
Single stars eject their envelope in an episode of strong winds at the end of the asymptotic giant branch (AGB) phase, and become contracting white dwarfs possibly giving rise to a planetary nebulae (PNe).
In the case of the presence of a close companion, the common envelope evolution leads to envelope ejection terminating the AGB phase or even the red giant branch (RGB) phase \citep{Kamath2015,Kamath2016,Kamath2019}.
Those objects are referred to as post-AGB or post-RGB binaries \citep[][]{VanWinckel2003}.

While the dust envelope surrounding the red giant star and its intrinsic variability limit our ability to detect a companion, once the envelope is ejected \citep[in the post-AGB or post-RGB phase;][]{VanWinckel2003,VanWinckel2018} binarity can be detected by radial velocity monitoring.
We can therefore observe the products of the strong binary interaction phase such as the resulting orbits and the properties of the circumbinary environment.
\modif{A disk-like infrared excess in the spectral energy distribution (SED) is strongly correlated with the presence of a binary system \citep{VanWinckel2018}.}

\modif{Spatially resolved observations of those cirucmbinary environments revealed their high degree of complexity} \citep{Hillen2014,Hillen2015,Kluska2018,Kluska2019} \modif{. O}nly intensive observing campaigns dedicated to imaging can uncover \modif{it} \citep{Hillen2016}.
The picture that emerges of these systems, based mainly on infrared interferometric imaging of \object{IRAS\,08544-4431} \citep{Hillen2016}, consists in a post-AGB star, a secondary with a circum-secondary accretion disk and a jet, a circumbinary disk perturbed by the inner binary, and extended emission of unknown origin (disk, wind, or jet?).
This picture needs to be confirmed with additional observations of other targets.

In this paper we present near-infrared interferometric observations of the evolved \modif{object} \object{HD101584}.
The main star's effective temperature was estimated to be around 8500\,K \citep{Sivarani1999,Kipper2005}.
The binarity of \object{HD101584} was \modif{inferred from} photometric and spectroscopic variations (Balmer jump, He\,I and C\,II) with a period of 218$\pm$0.7days \citep{Bakker1996}.
However, a period of 144\,days was also claimed from spectroscopic observations \citep{Diaz2007}.
\modif{Hence, there exists considerable uncertainty in the binary characteristics, even to the extent that the existence of a present companion can be discussed. 
However, the large-scale structure of the circumstellar environment seen in the Atacama Large Millimeter/submillimeter Array (ALMA) data (see below) strongly suggests that \object{HD101584} was a binary that went through a common-envelope process. 
Whether this ended before or by a merger plays no role for the interpretation in this paper, and we will therefore refer to the source as the \object{HD101584} system. 
In Sect.\,\ref{sec:separation}, we will discuss the implication of our observations on the possible present binary nature of \object{HD101584}.}

The Gaia \modif{parallax of} \object{HD101584}, 0.48$\pm$0.04\,mas, is of the order of the \modif{possible} binary separation (as derived in Sect.\,\ref{sec:separation}). Therefore, the distance estimate \modif{may be} biased by the orbital movement of the primary. 
Following \citet{Olofsson2019}, we give the distance-dependent results as their values at 1 kpc, as well as their scaling with distance.
Despite the lack of a reliable distance to \object{HD101584}, the evolutionary status of this system was determined to be, most likely, a post-RGB binary \citep{Olofsson2019}.
Its low $^\mathrm{12}$CO/$^\mathrm{13}$CO ratio, $\sim$13, indicates that it is on or beyond the RGB.
The $^\mathrm{17}$O/$^\mathrm{18}$O ratio of 0.2\modif{$\pm0.08$} points towards a low initial mass of the star \citep[$\lesssim$1.\modiff{3 }M$_\odot$ assuming an initial solar abundance ratio of 0.18;][]{Karakas2014,deNutte2017}. 
The higher luminosity of a post-AGB star would require a larger distance to the system \modif{(that would actually match the Gaia parallax)} than in the case of a post-RGB star.
\modif{This would make the estimated initial mass (the present stellar mass plus the mass of the ejected gas) less consistent with the initial stellar mass estimated from the oxygen isotope ratios.}
Hence, a post-RGB evolutionary status is preferred.

This system was extensively studied with \modif{the Atacama Large Millimeter/Submillimeter Array} \citep[ALMA;][]{Olofsson2015,Olofsson2017,Olofsson2019}.
The morphology of the \modif{stellar} environment as traced by sub-mm gas and dust emission can be separated into four components: the central compact source (CCS), the equatorial density enhancement (EDE), the high-velocity outflow (HVO) and the hourglass structure (HGS).
The CCS has a full width of half maximum size smaller than 100\,[D/1kpc]\,au\footnote{As mentioned, throughout the paper we give quantities for a reference distance of 1 kpc followed by the scaling of these quantities with distance given in squared brackets.} and emits in both continuum and numerous molecular lines.
It coincides in size with the mid-infrared emission detected by interferometry \citep{Hillen2017}.
The most likely interpretation is that the CCS corresponds to the circum\modif{stellar} disk.
The EDE is seen mainly in molecular lines, is in slow expansion and has an almost circular structure of a diameter of $\sim$2". 
It is interpreted as an almost face-on disk-like (or torus-like) structure with a possible connection to the CCS.
The HVO is elongated in the direction of PA$\sim$90$^\circ$.
It extends to a projected distance of 4" from the CCS in both directions, reaching velocities of 150\,km/s.
The blue shifted side points to the west.
The HVO ends at extreme-velocity spots (at $\pm$150\,km/s with respect to the systemic velocity) seen on both sides of the CCS in many molecular lines.
The HVO and the extreme-velocity spots reveal a slightly S-shaped morphology possibly indicating jet precession.
Finally, the HGS traced in CO(2-1) has a roughly elliptical morphology and surrounds the HVO.
It is likely produced by pressure from the HVO towards its sides.
There is also evidence for a secondary hourglass structure, that has a slightly different orientation, possibly tracing a previous outflow event.

\modif{It is interesting to note that nebulae with such extreme circumstellar characteristics are also seen around objects, claimed to be stellar mergers. 
For example, \object{CK Vul}, that was recently observed by ALMA, has a similar bipolar nebula and high outflow velocities  \citep[$\sim$100\,km/s;][]{Eyres2018}. The nebula is associated with a Luminous Red Nova (LRN) event that took place in 1670 \citep{Shara1985}.
It is advocated that the cause of an LRN may either be a giant eruption or, more likely, a stellar merger \citep{Pastorello2019}. 
The latter was argued in the case of \object{CK\,Vul} \citep{Kato2003,Eyres2018}. 
The luminosity of the central star of \object{CK Vul} is low however, around 1\,L$_\odot$ and its infrared excess does not show a continuous disk-like excess SED characteristic as in the cases of \object{HD101584} and post-AGB binaries. 
Nevertheless, another target, the Boomerang Nebula, with a more luminous central star (around 100\,L$_\odot$) and a disk-like SED, also has a similar large scale morphology. It was argued to be a post-AGB or post-RGB merger based on energy balance computations \citep{Sahai2017}. }

In this paper we present high-angular resolution observations in the near-infrared to uncover the structure of the inner environment of \object{HD101584} and probe the launching region of the bipolar jet.
We also aim to put additional constraint on the evolutionary status by resolving the inner binary and obtain the angular separation.
Because the object has a complex structure, we performed image reconstruction to retrieve the emission morphology independently of a model.
We first present the infrared interferometric observations (Sect.\,\ref{sec:obs}) and describe the image reconstruction (Sect.\,\ref{sec:ImgRec}). 
Then we applied a geometrical model to retrieve geometrical and thermal information on the different observed components of the system (Sect.\,\ref{sec:Modelfit}).
Finally, we discuss our findings and interpret the observed morphology (Sect.\,\ref{sec:dis}) before presenting our conclusions (Sect.\,\ref{sec:ccl}).


\section{Observations}
\label{sec:obs}

 \begin{figure}
 \centering
 \includegraphics[width=9cm]{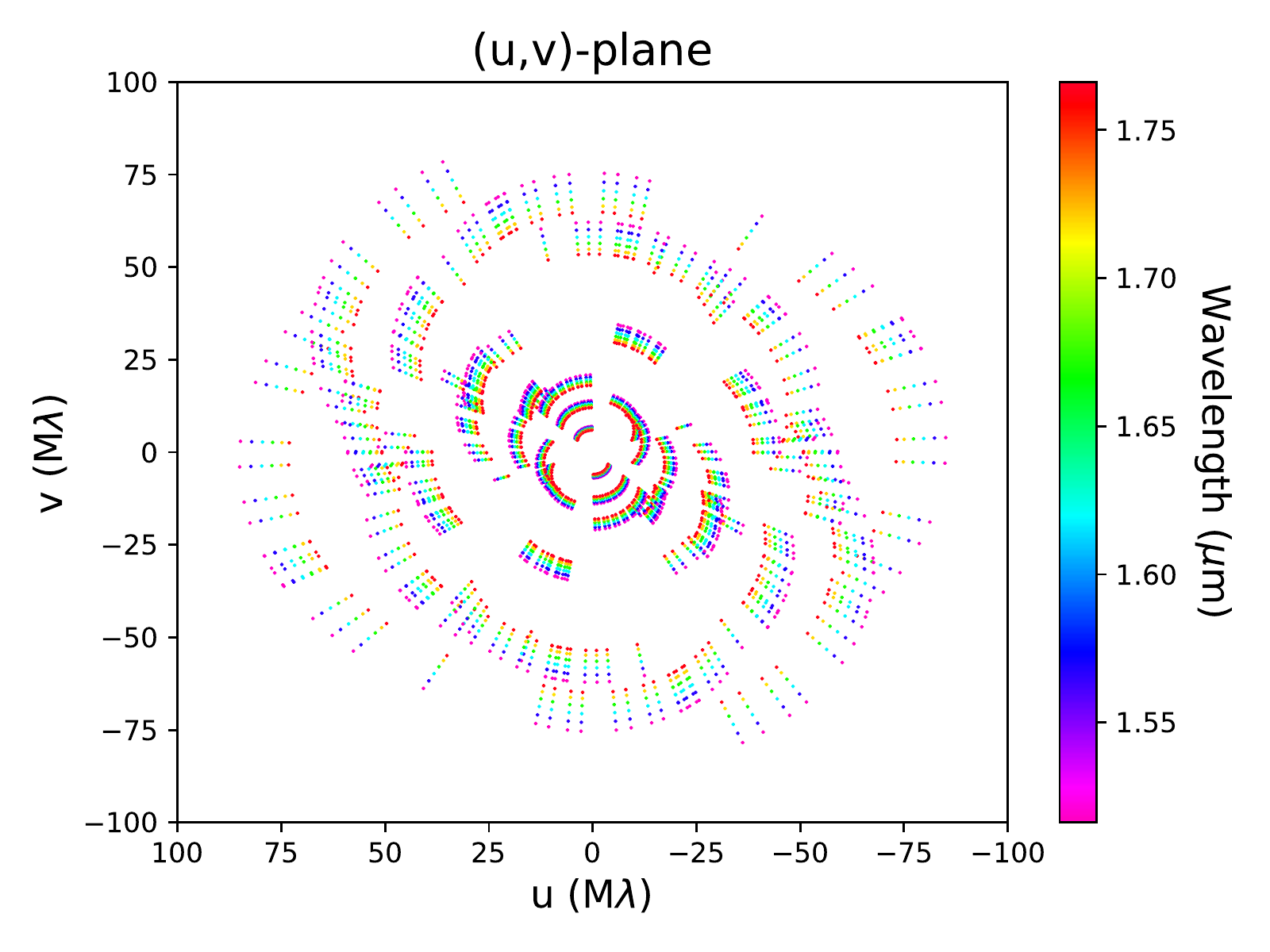}
  \includegraphics[width=9cm]{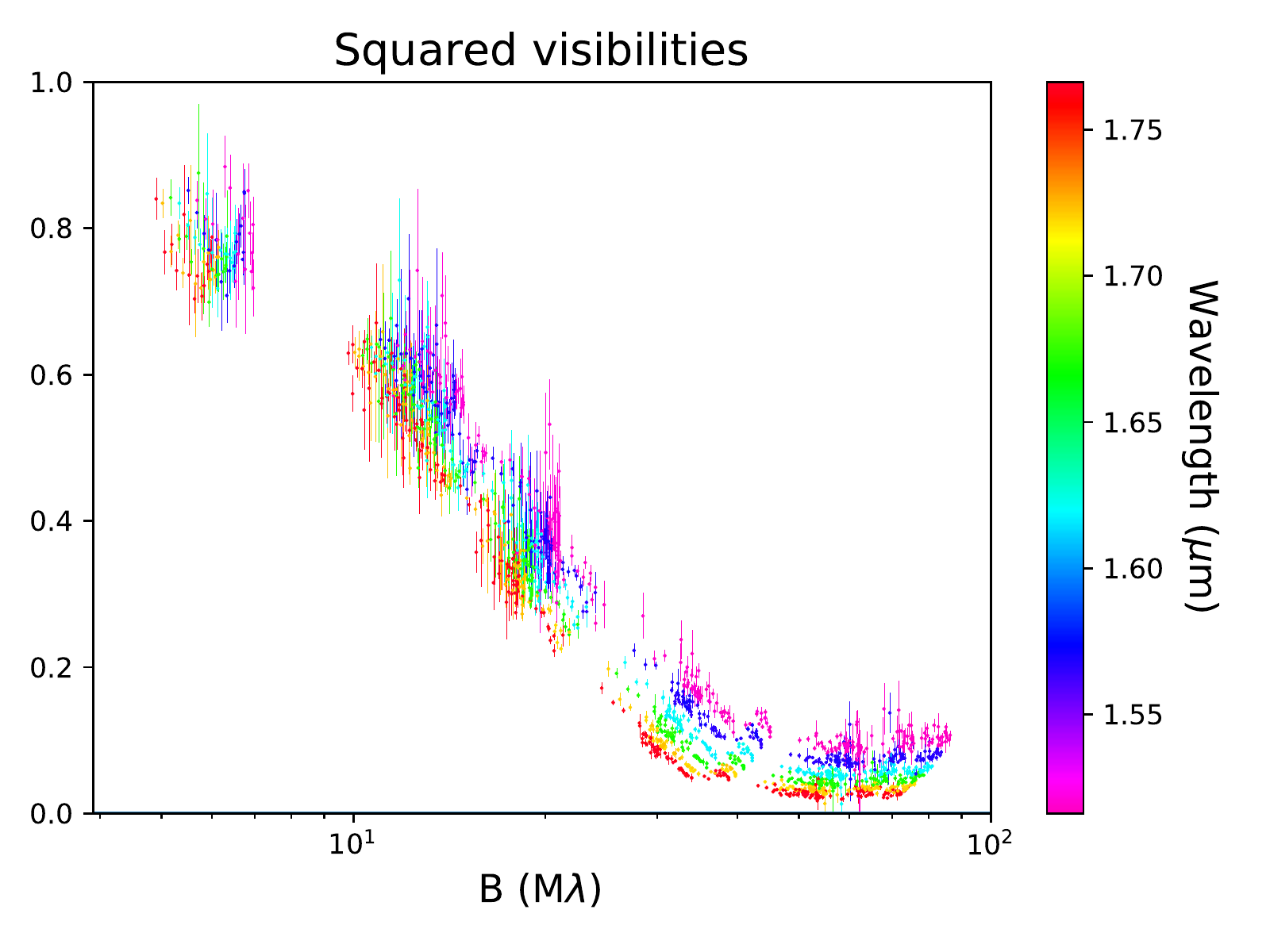}
    \includegraphics[width=9cm]{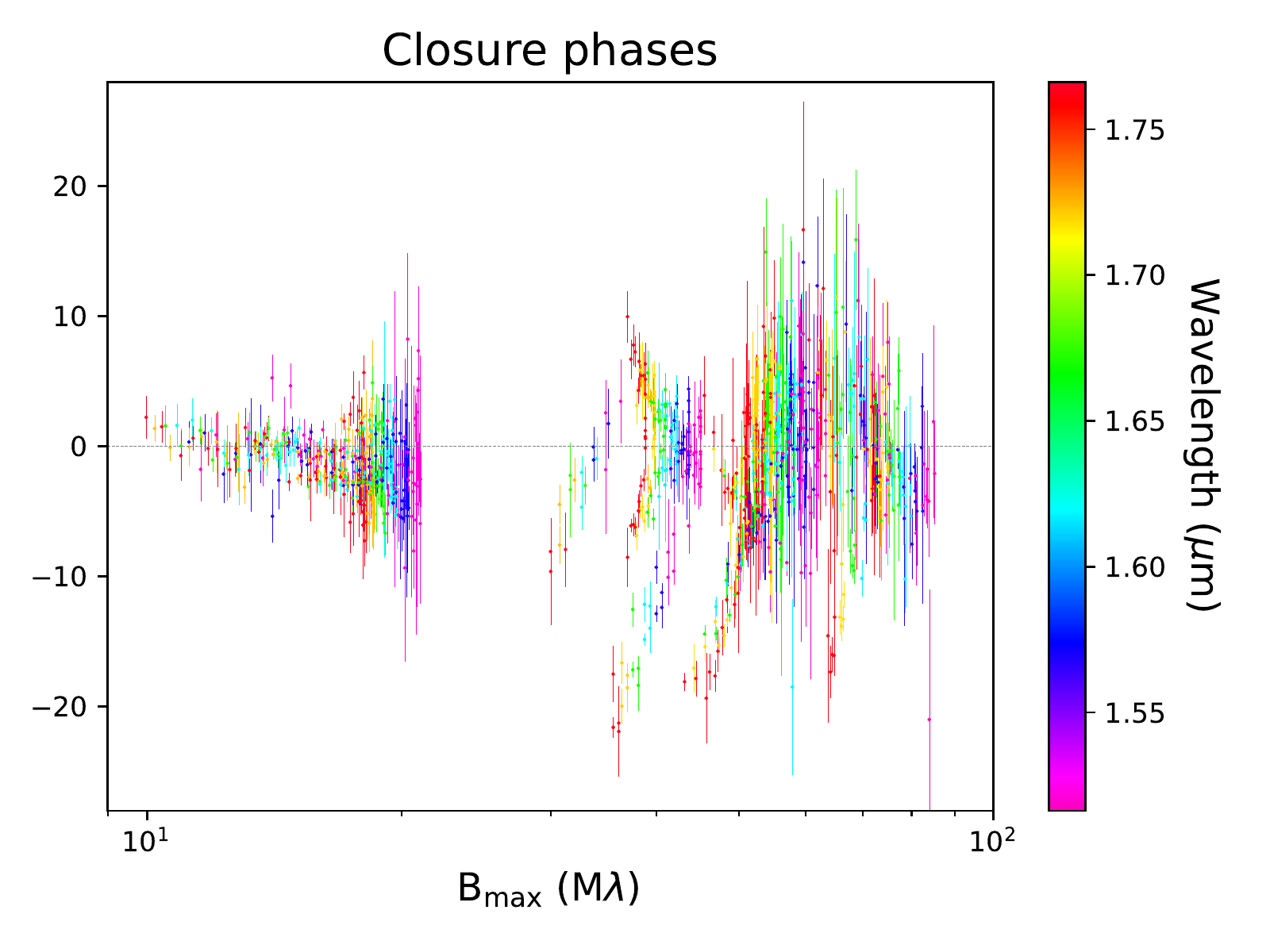}
 \caption{VLTI/PIONIER dataset on \object{HD101584}. Top: the obtained (u, v)-coverage. Center: the $V^2$ against spatial frequency and bottom: CP against the highest spatial frequency of a closure triangle. Spatial frequency is in log-scale. The color indicates the wavelength.}
 \label{fig:data}%
 \end{figure}
%
   

The observations were done using the Precision Integrated-Optics Near-infrared Imaging ExpeRiment (PIONIER) instrument \citep{LeBouquin2011} at the Very Large Telescope Interferometer (VLTI) between 2017-04-10 and 2017-05-20 (see Table\,\ref{tab:log}).
PIONIER is a four-beam interferometric combiner in the near-infrared $H$-band (central wavelength of 1.65$\mu$m).
The $H$-band is covered with 6 wavelength channels (R$\sim$40).
This instrument provides two observables: the squared visibilities (V$^2$) that are related to the size of the target and closure phases (CP) that are related to the degree of (non-)point-symmetry.
The data is plotted in Fig.\,\ref{fig:data} and is made of 297 V$^2$ and 194 CP measurements.
The (u, v)-plane is relatively well covered in every direction covering scales from 1 to 21\,mas (defined as $\lambda/2B$, where $\lambda$ is the observed wavelength and $B$ is the baseline).
From the V$^2$ we can see that their level decreases sharply with spatial frequency (defined as the ratio between the baseline and the observational wavelength) meaning that the observed object is spatially resolved.
\modif{At zero baseline V$^2$ reaches unity. 
However, for the shortest baselines, V$^2$ reach only 0.8 level, suggesting the presence of over-resolved emission.}
The V$^2$ do not reach zero at long baselines meaning that part of the total flux is unresolved.
The data is chromatic in such a way that V$^2$ for short wavelengths are at higher levels than for long ones.
It can be interpreted as the presence of a colder resolved component and an unresolved hot component as seen in other targets with circumstellar material \citep[e.g., protoplanetary disks around young stars;][]{Kluska2014}.
Some CPs are significantly non-null meaning that the source is not point symmetric.

\section{Image reconstruction}
\label{sec:ImgRec}

To interpret the V$^2$ and CP in a model-independent way, we use the technique of image reconstruction as the (u, v)-coverage is good enough for an image to be reliably reconstructed.
This technique enables determining the brightness distribution of the source model-independently to reveal any unpredicted morphologies.
In this paper, we will use the Multi-aperture Image Reconstruction Algorithm (MiRA) algorithm \citep{Thiebaut2008} together with the Semi-Parametric Approach to Reconstruct Chromatic Objects \citep[SPARCO;][]{Kluska2014}.
This approach allows one component of the target to be modelled and the other to be reconstructed by taking into account their temperature difference.
For example, the star can be modelled as a point source and its environment is reconstructed by taking into account the spectral index of each of the components.
Given the angular scales probed by the (u, v)-coverage, our image is set to have 256$\times$256 pixels with a pixel size of 0.15\,mas.
To reconstruct the image, the algorithm is minimizing a cost function ($f$), defined from the Bayesian equation, that consists of two terms: $f(x) = f_\mathrm{data}(x) + \mu f_\mathrm{rgl}(x)$, where $x$ is the image, $f_\mathrm{data}$ is the data likelihood term (here the $\chi^2$), $f_\mathrm{rgl}$ is the regularization term and $\mu$ the regularization weight.
We used the quadratic smoothness regularization that is defined as: $f_\mathrm{rgl} = x - S.x$, where $S$ is a smoothing operator.
This regularization was proven to be one of the best to be used in infrared interferometry \citep{Renard2011} and is performing well in practice \citep[e.g.][]{Kluska2016,Hillen2016}.

 \begin{figure}
 \centering
 \includegraphics[width=9cm]{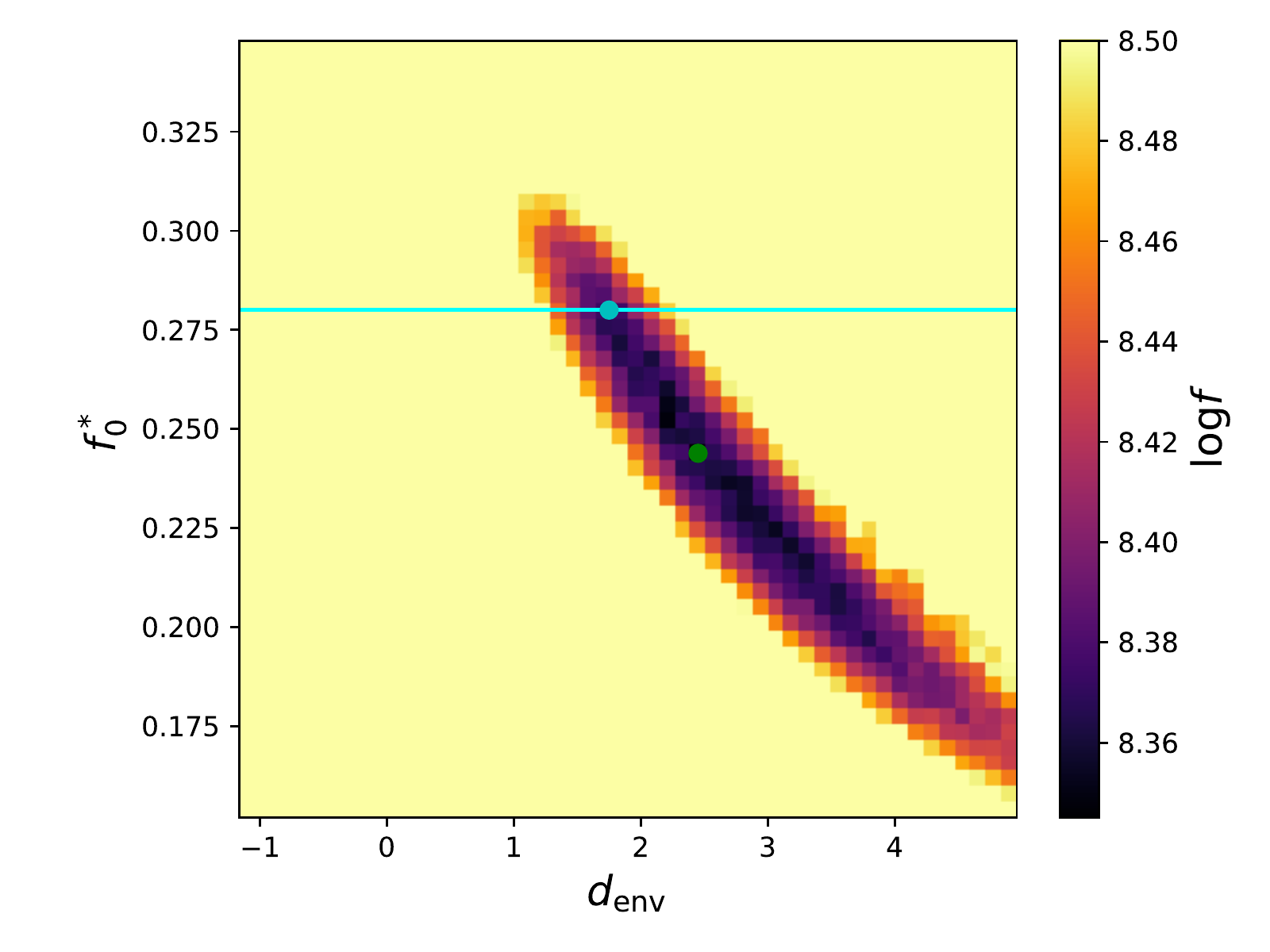}
 \caption{Logarithm of the cost function $f$ as function of the chromatic parameters \fso and \denv. The green circle represents the pair of chromatic parameters where the cost-function value is the smallest. The cyan horizontal line is the stellar-to-total flux ratio from the best fit geometrical model (\fso=0.28). The cyan circle is the pair of chromatic parameters giving the most likely image reconstruction using \fso from the geometrical model fitting.}
 \label{fig:chrom}%
 \end{figure}

To reconstruct the image with the SPARCO approach one needs to determine the regularization weight ($\mu$) and the chromatic parameters, i.e., the stellar-to-total flux ratio at 1.65\,$\mu$m (\fso) and the spectral index of the environment (\denv).
This process is described in detail in \citet{Kluska2016}, but we recall the steps here: we choose $\mu$ and perform a grid on \fso and \denv to find the pair of these parameters that minimizes the cost function $f$ (see Appendix\,\ref{app:chrom}).
For this object the chromatic parameters are degenerate (see Fig.\,\ref{fig:chrom}).
We nevertheless chose the pair of chromatic parameters with the lower $f$ to obtain the final image.
However, in Sect.\,\ref{sec:Modelfit} we perform an image reconstruction with another pair of chromatic parameters constrained from our geometric model, showing the impact of different chromatic parameters on the image.
Finally, to determine the significant structure in the image we performed a bootstrap on the dataset.
It consists in building a new dataset by drawing baselines for V$^2$ and closure triangles for CP until we reach the same number of data points.
Each baseline or triangle can be drawn several times or none.
In order to find the optical regularization weight we have used the L-curve method as described in \citep{Kluska2016}. 
For different regularization weights $\mu$ we compute both the values of the likelihood ($f_\mathrm{data}$) and the regularization terms ($f_\mathrm{rgl}$) values. 
On a log-log plot those values will trace two asymptotes for low and high values of the regularization weight tracing the regimes where the likelihood and the regularization term dominates.
Finally, we have chosen to use $\mu=10^9$ where the image is well regularized while still reproducing the data (we detail the impact of the regularization in Sect.\,\ref{sec:mu}). 

 \begin{figure*}[!th]
 \centering
\includegraphics[width=9cm]{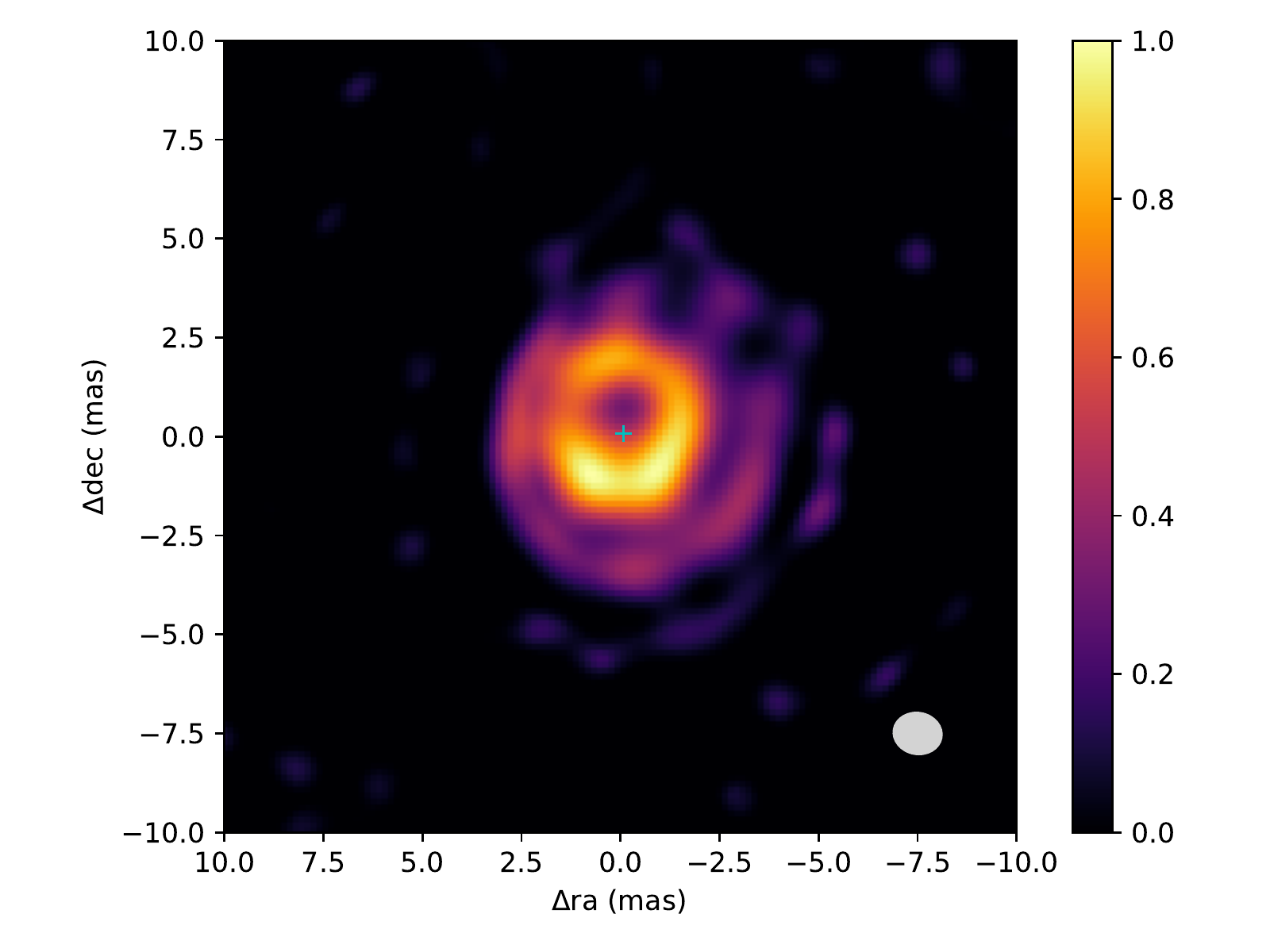}
 \includegraphics[width=9cm]{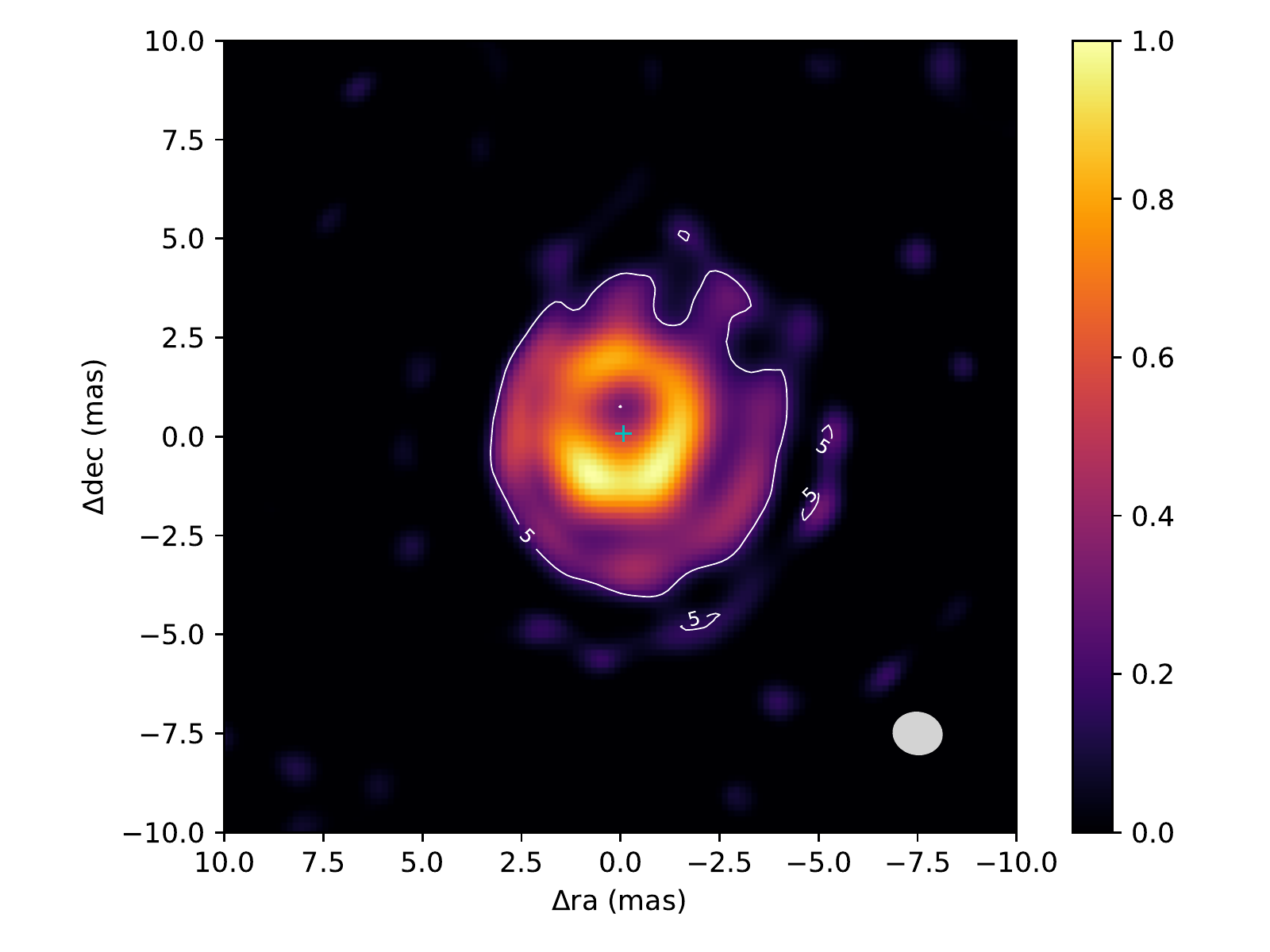}
 \includegraphics[width=9cm]{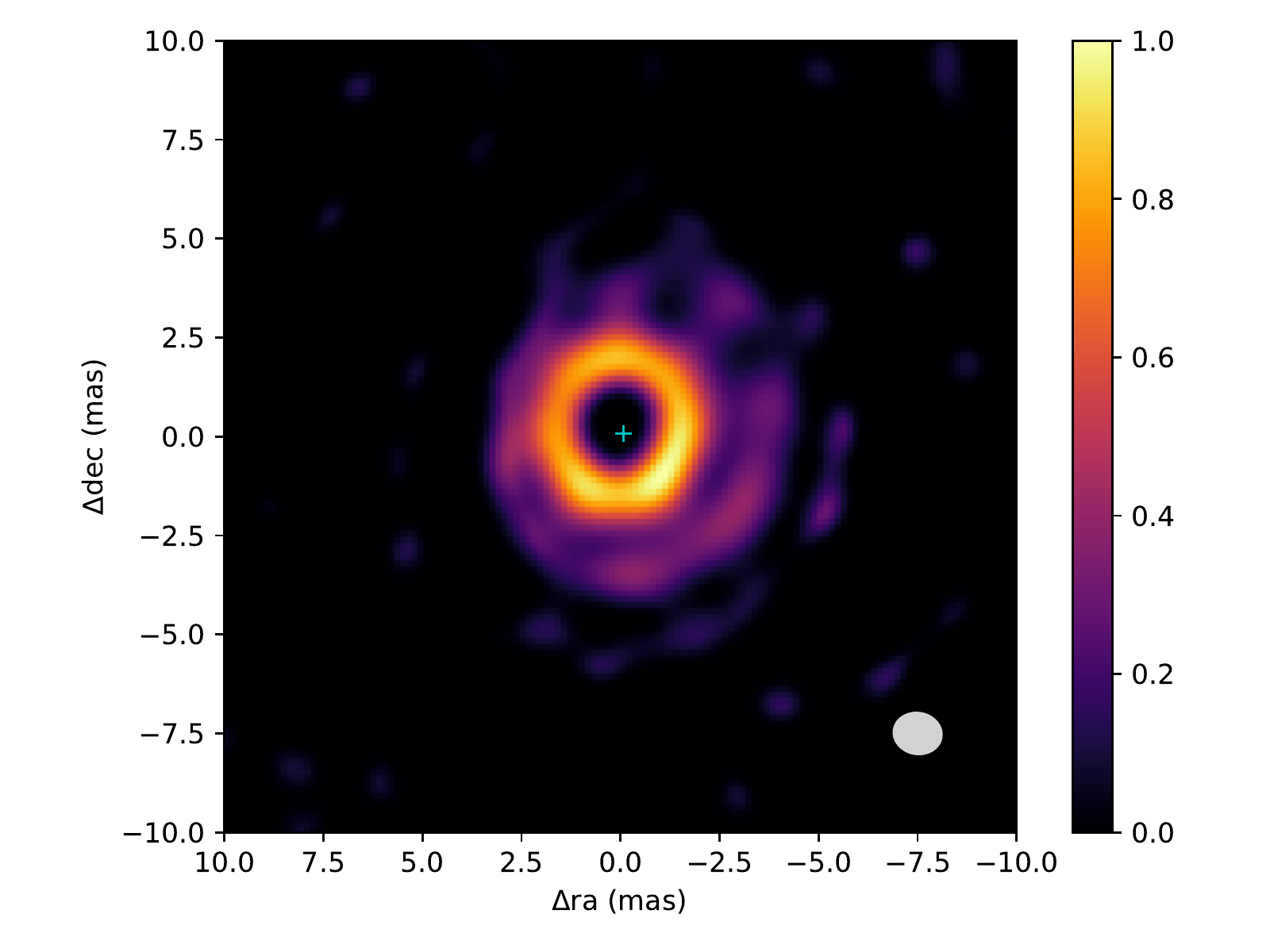}
 \includegraphics[width=9cm]{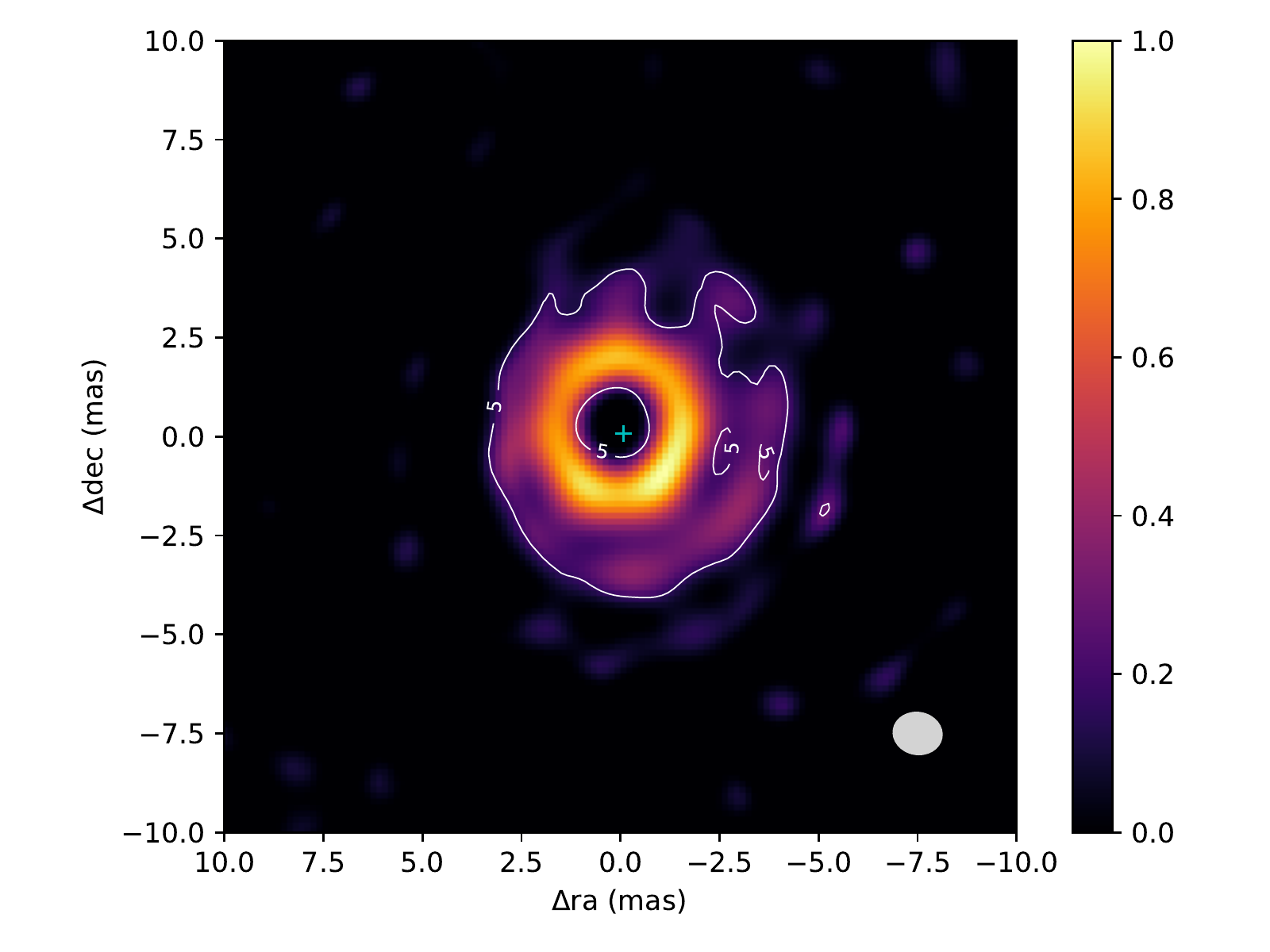}
 \caption{Image reconstruction of \object{HD101584} with\modif{out} (\modif{left}) and with (\modif{right}) the significance contours \modif{at 5-$\sigma$ from bootstrapping}. \modif{Top images for \fso=0.244 and \denv= 2.45. Bottom images for \fso=0.28 and \denv=-1.75}. The position of the star (which is not in the image) is indicated by the green cross. The white contours are the 5-$\sigma$ significance contours. The beam size is indicated in the lower-right corner \modif{of each image}.}
 \label{fig:image}%
 \end{figure*}


In Fig.\,\ref{fig:image} we can see the image reconstruction with the $5$-$\sigma$ significance contour.
The image reconstruction reproduces the data well with a reduced $\chi^2_\mathrm{red}$ of 1.38 (see also Fig.\,\ref{fig:Image1vsdata}).
The best pair of chromatic parameters are \fso= 0.244 and \denv= 2.45.
This translates into a stellar-to-total flux ratio of 24.4\% at 1.65$\mu$m and a\modif{n image} spectral index of a black body having a temperature of $\sim$1170\,K assuming a spectral index of $\mathrm{d}\log F_\lambda/\mathrm{d}\log \lambda=-3.1$ \citep[as expected for a T$_\mathrm{eff}$$\simeq$8500\,K;][]{Olofsson2019} for the central \modiff{star}.
In the image (Fig.\,\ref{fig:image}) there is a complex structure that appears to be two rings.
The brightest one being close to the star and the fainter one being larger and probably slightly shifted w.r.t. the first one.
The two rings are above $5$-$\sigma$ significance in the image.
There are some small features that are marginally more significant that $5$-$\sigma$, but, as their existence is not certain, we do not take them into account in this study.

\section{Geometrical modeling}
\label{sec:Modelfit}

Having established the morphological features present in the image (a hot point source and two colder rings), we want to obtain quantitative values describing the morphology and the relative emissions of the different components of the system.
To achieve this we perform geometrical model fitting.
Inspired by the image reconstruction we designed a geometrical model that is composed of a point source to model the central star, two rings, and a background for the over-resolved flux.
The model is described in Appendix\,\ref{app:model} and has 15 parameters.
The central star is defined by three parameters: the stellar-to-total flux ratio ($f^*_0$) and its coordinates relative to the center of the inner ring ($\Delta x$ and $\Delta y$).
The rings are defined by their diameters ($rD_1$ and $rD_2$ for the inner and outer ring, respectively), their full width at half maximum in units of the ring radius ($rW_1$ and $rW_2$, respectively), their ring-to-total flux ratios ($f^\mathrm{ring1}_0$ and $f^\mathrm{ring2}_0$), and their black body temperatures ($T_1$ and $T_2$) constrained from the changing flux ratios between the components using all spectral channels.
We assume that one of the rings belongs to a disk called the mid plane. 
The other ring could be in the same plane (and part of the same disk), or it can be located out of this plane and part of another structure, e.g., the hourglass structure (HGS) as discussed below.
We assume their inclinations ($i$) and position angles (PA) to be the same, but we allow the outer ring to be shifted w.r.t to the inner ring in the direction of the minor-axis ($\Delta R_\mathrm{ring2}$).
Finally, the back-ground is defined by the background-to-total flux ratio ($f^\mathrm{bg}_0$) and its spectral index ($d_\mathrm{bg}$).
Because the flux ratios are relative quantities the ring1-to-total flux ratio is not fitted but deduced from the others as being $f^\mathrm{ring1}_0$ = 1-$f^\mathrm{*}_0$-$f^\mathrm{ring2}_0$-$f^\mathrm{bg}_0$.
In order to explore the parameters of such a complex model we have used a two-steps approach already used and explained in \citet{Kluska2019}.
The geometrical model is transformed into V$^2$ and CP values for each baseline of the original dataset.
The fit is then performed in the Fourier domain.
We start with a distribution of first guesses of the parameters based on the image reconstruction.
We then explore the parameter space with a genetic algorithm \citep[using Distributed Evolutionary Algoirthms in Python framework;][]{DEAP} that efficiently finds minima.
Then, we use the best solutions from the previous step with the Markov chain Monte Carlo sampler \texttt{emcee} \citep{MCMC} to derive errors.
\modiff{The errors on several parameters are very small (Fig.\,\ref{fig:MCMC}) very likely because the geometrical model, while having fifteen parameters, is still too simple to reproduce the whole dataset. There is therefore a possible model bias that results in small errors.}

\begin{figure}
\centering
\includegraphics[width=9cm]{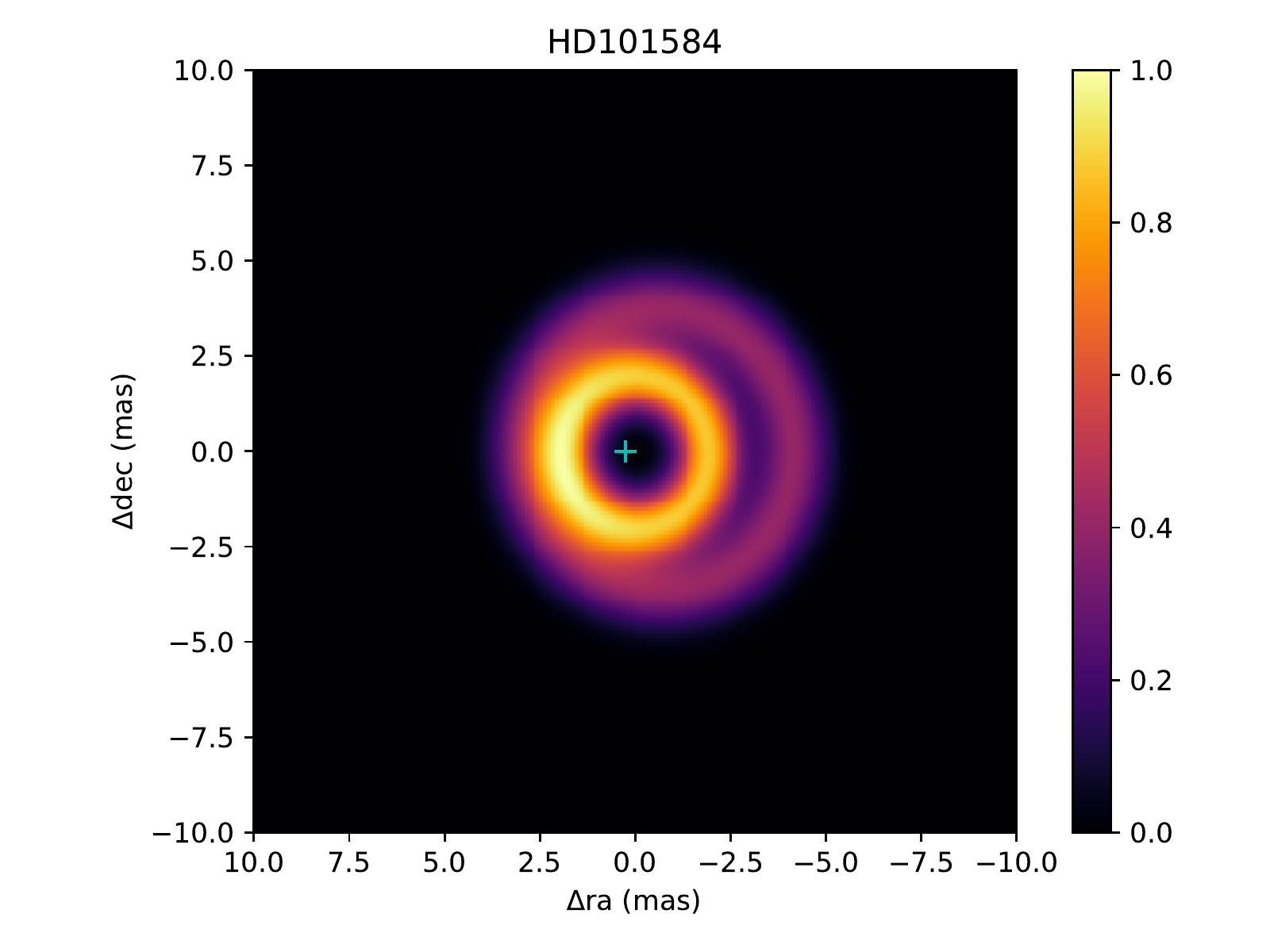}
\includegraphics[width=8cm]{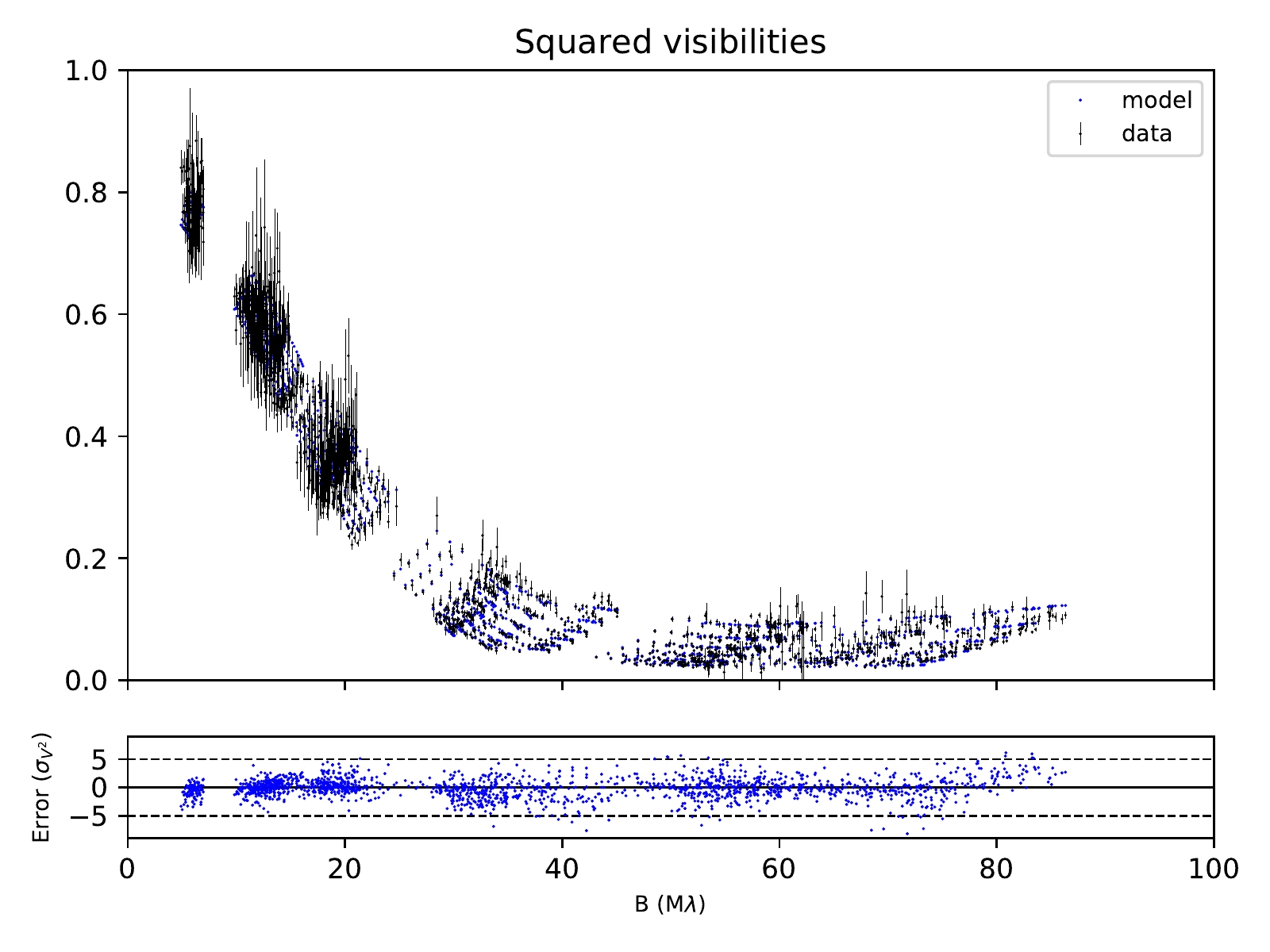}
\includegraphics[width=9cm]{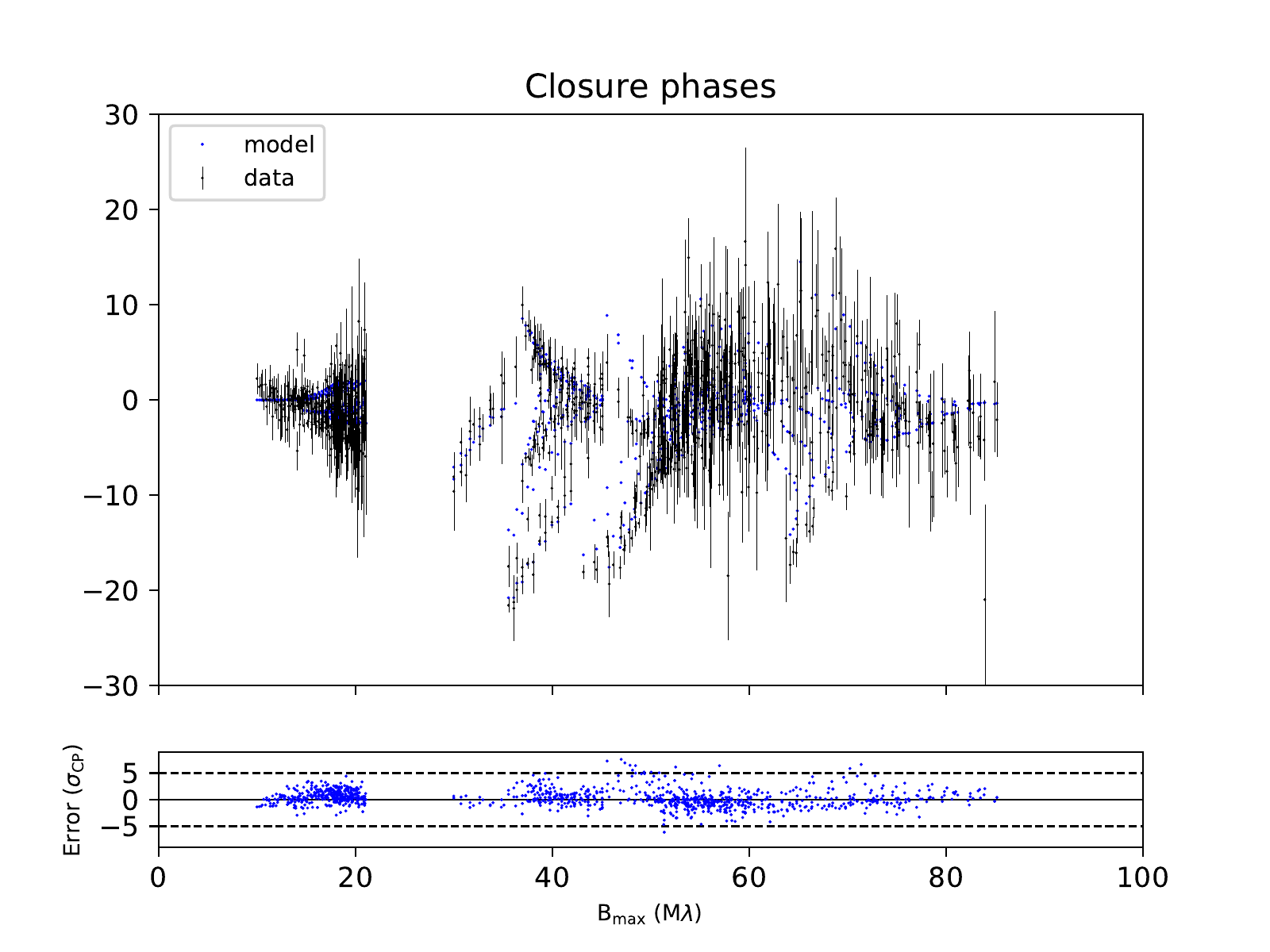}
\caption{Best-fit model of \object{HD101584}. Top: Image of the best-fit geometrical model. Comparison between the best-fit model and data V$^2$ (center) and CP (bottom).  }
\label{fig:imageparam}%
\end{figure}

\begin{table}
\caption{Best fit parameters of the geometrical model}             
\label{tab:fitres}      
\centering                          
\begin{tabular}{c c c c c c}        
\hline\hline     \\     [-10pt]        
 $\chi^2_\mathrm{r}$ & f$^*_0$& f$^\mathrm{ring1}_0$~ \tablefootmark{a}& f$^\mathrm{ring2}_0$& f$^\mathrm{bg}_0$  &  d$_\mathrm{bg}$  \\  
 & [\%] &[\%]  & [\%] & \\[3pt] 
\hline \\     [-10pt]
   3.1& $28.0^{0.1}_{-0.1}$& $42.0^{0.3}_{-0.3}$ & $20.9^{0.3}_{-0.3}$& $9.1^{0.1}_{-0.1}$ & $2.3^{0.2}_{-0.2}$\\ [3pt] 
 \hline
 \hline \\   [-10pt]
 &T$_1$&rD$_1$& rW$_1$\\
 & [K]  & [mas] &  \\[3pt] 
  \hline\\  [-10pt]
  &$1540^{10}_{-10}$&  $3.94^{0.01}_{-0.01}$& $0.75^{0.01}_{-0.01}$\\[3pt]
  \hline
 \hline \\   [-10pt]
  &T$_2$& rD$_2$& rW$_2$ & $\Delta$R$_\mathrm{ring2}$ \\
 & [K]  & [mas] &  & [mas] \\[3pt] 
    \hline\\  [-10pt]
 &  $1014^{10}_{-10}$&$7.39^{0.02}_{-0.02}$& $0.43^{0.01}_{-0.01}$& $0.57^{0.01}_{-0.01}$\\[3pt]
   \hline
 \hline \\   [-10pt]
 & $i$& PA  & $\Delta$x& $\Delta$y \\ [1pt]  
& [deg] & [deg] & [mas] & [mas] & \\ [0pt] 
\hline            \\         [-10pt]              
&$19.2^{0.4}_{-0.5}$& $8.6^{0.4}_{-0.4}$ & $0.26^{0.01}_{-0.01}$& $0.00^{0.01}_{-0.03}$  \\ [0pt] 
\end{tabular}
\tablefoot{\tablefoottext{a}{f$^\mathrm{ring1}_0$ is not fitted but deduced from the other flux ratios as $f^\mathrm{ring1}_0$ = 1-$f^\mathrm{*}_0$-$f^\mathrm{ring2}_0$-$f^\mathrm{bg}_0$.}}
\end{table}

The best fit parameters are summarized in Table\,\ref{tab:fitres} and the associated image is shown in Fig.\,\ref{fig:imageparam}.
The fit quality is decent as it reaches a reduced $\chi^2$ of 3.1 (with 15 degrees of freedom).
The image is similar to the reconstructed image, the two rings having similar sizes as the features in the reconstructed image.
However, in the geometric model, there is less flux close to the star than in the image reconstruction.
This is due to the structure of the rings in the geometrical model being defined to have a Gaussian profile.
As a result the stellar-to-total flux ratio (f$^*_0$) at 1.65$\mu$m is higher in the geometrical model (28.0$^{0.1}_{-0.1}$\% compared to 24.4\% in the image reconstruction).
The whole system has an inclination ($i$) of 19.2$^{+0.4}_{-0.5}$\,degrees with a position angle (PA) of 8.6$^{+0.4}_{-0.4}$\,degrees.

Hereafter, we will call ring1 the inner ring and ring2 the outer ring for clarity.
The inner ring has a diameter (rD$_1$) of 3.94$^{0.01}_{-0.01}$\,mas, a full width at half maximum of 1.48\,mas, and a temperature ($T_1$) of 1540$^{10}_{-10}$\,K.
The outer ring is almost two times larger than the inner one (rD$_2$ = 7.39$^{0.02}_{-0.02}$\,mas).
The center of the outer ring appears to be shifted by 0.57$^{0.01}_{-0.01}$\,mas compared to that of the inner ring (in the direction of the minor axis).
It is also colder than the inner ring ($T_2$=1014$^{10}_{-10}$\,K).
A non-negligible part of the circum-stellar flux is modelled as a background with a size larger than 20\,mas (f$^\mathrm{bg}_0$=$9.1^{0.1}_{-0.1}$\%) with a spectral index (d$_\mathrm{bg}$) of $2.3^{0.2}_{-0.2}$ (corresponding to a black body emission of $\sim$1200\,K).

Finally, because the chromatic parameters for the first image reconstruction were degenerate, we perform another image reconstruction using the stellar-to-total flux ratio ($f^*_0=0.28$) found in the geometrical fit.
We used the same parameters as in the first image reconstruction, and made a grid on the spectral index of the environment (\denv) only.
We found \denv=1.75$\pm$0.25 (corresponding to a black body temperature of $\sim$1290\,K) to be the most optimal value in accordance with the grid made for the first image reconstruction (see Fig.\,\ref{fig:chrom}).
We retrieve an image with very similar features, except that there is less emission close to the star as expected (Fig.\,\ref{fig:image}), confirming the double ring morphology.
This image reconstruction has a similar $\chi^2_\mathrm{red}$ as the previous one (1.39) and is reproducing the data equally well (see Fig.\,\ref{fig:Image1vsdata}).

\section{Discussion}
\label{sec:dis}

\subsection{Rings are from thermal emission}

\modif{First, we need to determine whether the origin of the rings emission is thermal or scattered light. 
Given the wavelength dependency of the visibilities in the $H$-band we can estimate the black body temperature associated with each ring. 
We found low temperatures (1540$\pm10$\,K for the inner ring and 1014$\pm$10\,K for the outer one) making it unlikely that scattered light dominates the emission as it would have a bluer spectrum similar to the stellar one. 
We note that 9.1$\pm$0.1\% of the flux is coming from an over-resolved component.
However, its spectral index indicates a low temperature (T$\sim$1290\,K) also not compatible with scattered stellar light.
}

\modif{The orientation (inclination and position angle) of the rings fits well the results from the large scale ALMA dataset \citep[HVO; 5$^\circ<i<20^\circ$; PA$\sim90^\circ$;][]{Olofsson2019}. 
The PA, 9$^\circ$, is essentially orthogonal to that of the HVO, suggesting an orientation of the rings in the equatorial plane of the outflow. 
We are, therefore, likely tracing the inner parts of the equatorial plane as seen with ALMA. 
}

\subsection{Rings in the same plane?}

\modif{The structures of the circumstellar environment, as resolved in this object, are complex. Here, we review the underlying mid-plane structures that could give such an image in the near-infrared.}

\modif{Both rings could be made by different mass loss episodes. 
Such a scenario was proposed to explain the double ring morphology in the equatorial plane of the Butterfly Nebula detected in $^{12}$CO\,J=2-1 and $^{12}$CO and $^{13}$CO \,J=3-2 and with the Institut de Radioastronomie Millim\'etrique (IRAM) Plateau de Bure interferometer and ALMA, respectively \citep{Castro2012,Castro2017}.
Its large-scale hourglass-shaped outflow is similar to the one of \object{HD101584}, but is seen edge-on.
However, the scale of the rings of \object{HD101584} (diameters of $\sim$7\,au and $\sim$13\,au for a distance of 1.8\,kpc) are very different to those of the Butterfly Nebula as those have diameters of $\sim$2600 and $\sim$5200\,au (for the adopted distance of 650\,pc).
It is interesting to note that the rings are offset by $\sim$130\,au.
Those rings are possibly related to the hourglass-shaped structures seen in the scattered light images from the HST \citep{Clyne2015}.
The Butterfly nebula has a lower estimated mass (nebular mass of $\sim10^{-4}\,M_\odot$) than the environment of \object{HD101584} by at least an order of magnitude (mass of the CCS only is estimated to be about $10^{-3}\,M_\odot$).
The different scales, and the absence of dust thermal emission in the Butterfly nebula make the link with the structure we observe weak.
Furthermore, we lack of kinematical information and cannot therefore conclude on a similar origin for the rings in the case of \object{HD101584}.
It is possible that at least one of the rings around \object{HD101584} is due to a mass loss event. 
This can be probed by re-imaging the target at an additional epoch in a search for dynamical changes.
}

\modif{\citet{Icke2003} argued that multi-polar circumstellar nebulae, as seem to be seen around \object{HD101584} \citep{Olofsson2019}, can arise from the interaction of the wind with a warped circumstellar or circumbinary disk structure. 
Our data do not show strong evidence for such warp as both rings are compatible with having the same orientation. }

\modif{
A binary system would likely induce a spiral wave in the circumbinary disk. 
From the images in Fig.\,\ref{fig:image}, one could argue that the outer ring could in fact be a spiral originating from the East side and developing through the South to the West.
This may in principle be possible if the disk is optically thin.
Detailed radiative transfer modeling of this dataset is beyond of the scope of this paper.
However, the radiative transfer model that was set to reproduce the SED, the ALMA continuum observations and the rough size from near-infrared interferometry \citep[the inner rim has a diameter of 4.3\,mas in this model;][]{Olofsson2019} shows that the inner disk region is dense enough to be radially optically thick, meaning that the rest of the disk is shadowed from the star by the inner disk.
It results in a sharp temperature drop behind the rim.
Such a result is also recovered when modeling circumbinary disks around post-AGB objects from near-infrared interferometric data \citep{Hillen2015,Kluska2018}.
Any structure behind the rim would not be detected in $H$-band.
A more refined model taking into account the more complex morphology revealed in this work would need to be constructed to further investigate this hypothesis.
}

\subsection{A disk inner rim and an out-of-plane ring?}
\label{sec:scenarios}

\begin{figure}
\centering
\includegraphics[width=9cm]{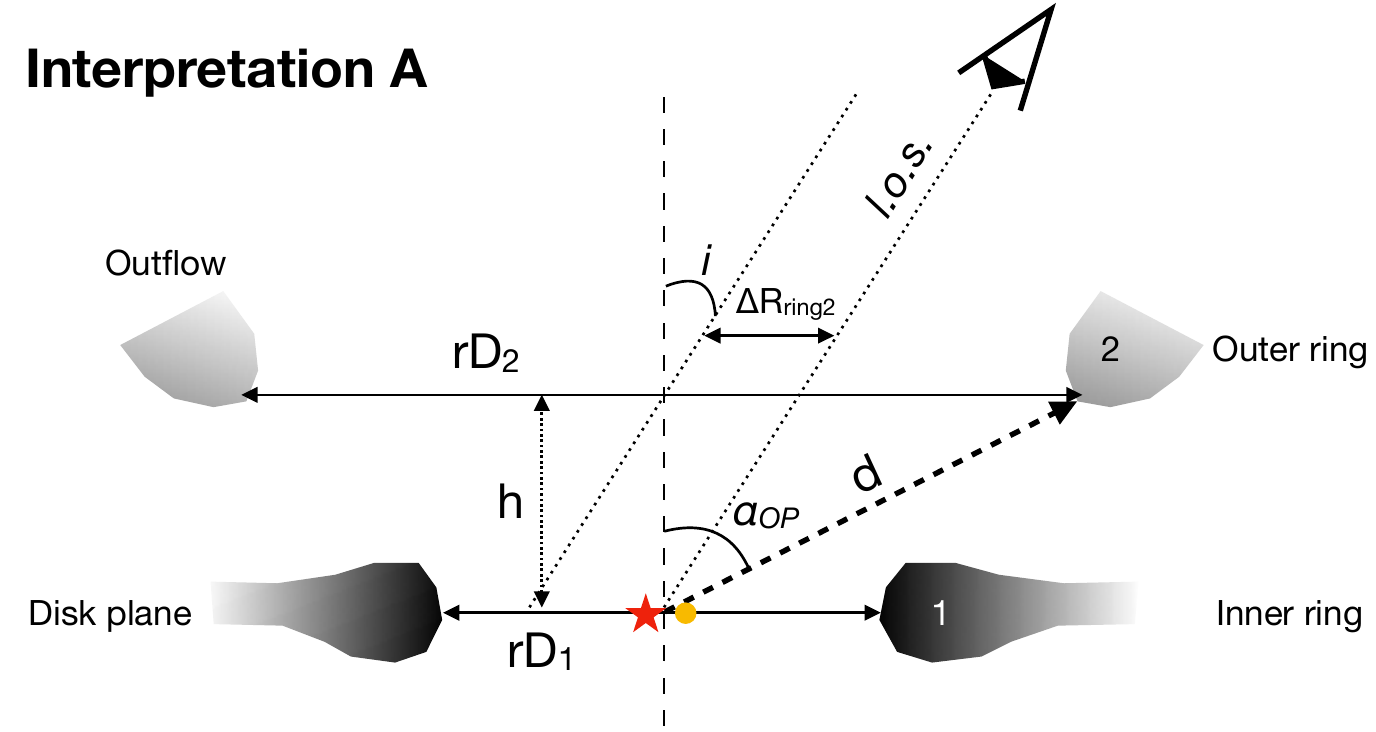}
\includegraphics[width=9cm]{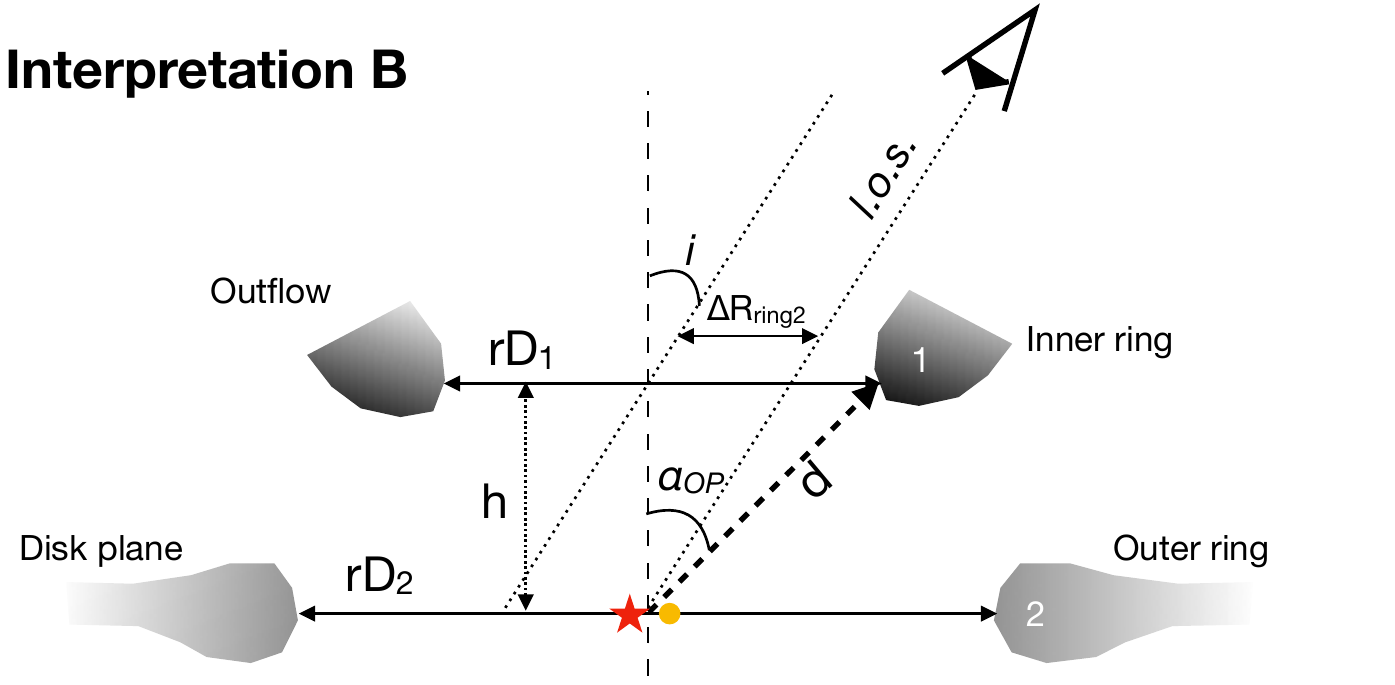}
\caption{Sketch of the two proposed interpretations. The evolved star is indicated by the red star and the unseen companion by the yellow disk.}
\label{fig:sketch}%
\end{figure}

\modif{An alternative interpretation is that one of the rings is not in the midplane.
Such a geometry may explain the} shift between the two rings \modif{as observed in the image}.
\modif{The ring in the midplane would then trace the disk inner rim that is likely ruled by dust sublimation \citep{Kluska2019}.}
As we do not have unambiguous information on which of the rings is above the mid-plane we will consider two interpretations. 
Interpretation A treats the case where the inner disk is in the mid-plane whereas interpretation B treats the case of the outer ring being in the mid-plane (see Fig.\,\ref{fig:sketch}).
\modif{For each of the interpretations we work out the three-dimensional configuration and the temperature of the dust located at different distances from the stars.}

\subsubsection{Interpretation A: Inner ring in the mid-plane}
\label{sec:ScA}

\begin{figure}
\centering
\includegraphics[width=9cm]{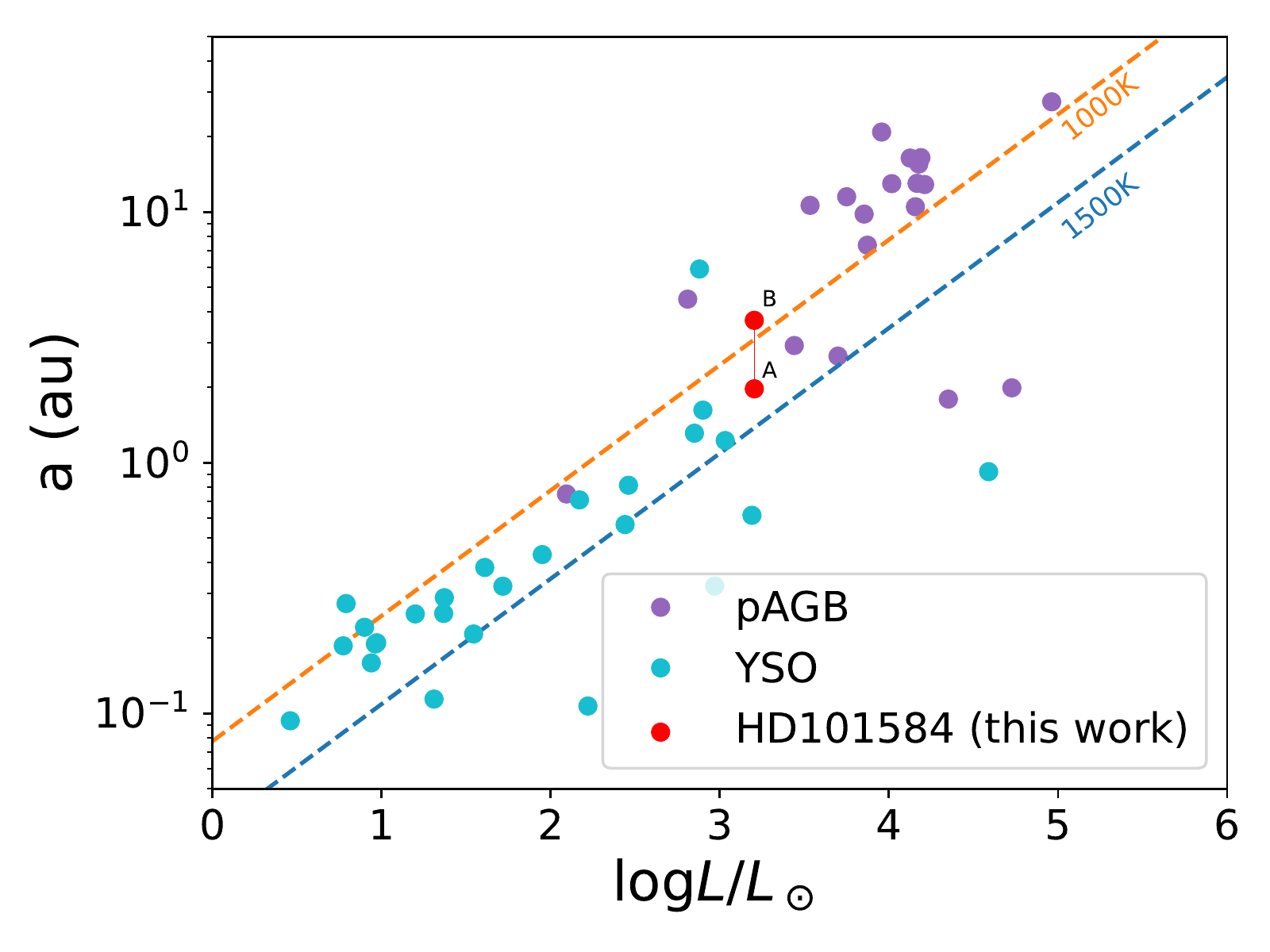}
\caption{Locating \object{HD101584} on the size-luminosity diagram from \citet{Kluska2019} for the two interpretations discussed in Sect.\,\ref{sec:scenarios}. }
\label{fig:Rsub}%
\end{figure}

In this interpretation, we assume that the inner ring is the disk inner rim that would be at the dust sublimation radius as in most of the similar post-AGB binary targets \citep{Kluska2019}.
Knowing the stellar luminosity one can use the following equation to determine the dust sublimation radius \citep[][]{Monnier2002,Lazareff2017}:
\begin{equation}
\label{eq:Tsub}
    a = \frac{1}{2} (C_\mathrm{bw}/\epsilon)^{1/2} (L_\mathrm{bol}/4\pi\sigma T^4_\mathrm{sub})^{1/2},
\end{equation}
where $T_\mathrm{sub}$ is the sublimation temperature, $C_\mathrm{bw}$ is the back-warming coefficient \citep[][]{kama2009}, $\epsilon$\,=\,$Q_\mathrm{abs}(T_\mathrm{sub})/Q_\mathrm{abs}(T_*)$ is the dust grain cooling efficiency which is the ratio of Planck-averaged absorption cross-sections at the dust sublimation and stellar temperatures, and $\sigma$ the Stefan-Boltzmann constant.
The luminosity of the central star was determined to be 1600\,L$_\odot$ \citep[assuming a distance of D=1\,kpc;][]{Olofsson2019}.
Assuming $Q_\mathrm{abs}(T_\mathrm{sub})/Q_\mathrm{abs}(T_*)=1$ and the back-warming coefficient being $C_\mathrm{bw}$=1, the theoretical dust sublimation radius would be 2.5 or 1.4\,[D/1kpc]\,au for a dust sublimation temperature of 1000\,K and 1500\,K, respectively that are  roughly the near-infrared emission temperatures found in post-AGB binaries \citep{Kluska2019} and around young stars \citep{Lazareff2017} respectively.
The measured inner rim radius\modif{,} 2.0\,[D/1kpc]\,au\modif{,} falls between the above estimates.
While this is a standard size for inner rims of protoplanetary disks, for most post-AGB circumbinary disks (similar to the object of this study) the size is usually slightly larger (see Fig.\,\ref{fig:Rsub}).
Nevertheless, \object{HD101584} does not stand out, and we can assume that the inner disk rim is ruled by dust sublimation physics.

The outer ring has a temperature that is lower than that of the inner ring.
We interpret the shift of the outer ring w.r.t. the inner ring as a geometrical effect due to the system inclination (see Fig.\,\ref{fig:sketch}).
One can therefore compute the height of the outer ring above the midplane ($h$) as being $h=\Delta$R$_\mathrm{ring2}/\tan i $.
It results in a height of 1.6\,[D/1kpc]\,au.
We can then verify the thermal emission origin of the outer ring by computing its distance to the central star, assuming it is located at the center of the inner ring.
This distance is $d=\sqrt{(rD_2/2)^2 + h^2}$, which gives 4.1\,[D/1kpc]\,au.
If we assume that the emission of both rings comes from thermal emission of dust with similar characteristics (opacity, density) we can predict the distance to the outer ring knowing its temperature as $T_1^2 rD_1/2 = d^\mathrm{theo.} T_2^2$.
In that case, $d^\mathrm{theo.} = 4.5$\,[D/1kpc]\,au, which is reasonably close to the estimated geometrical distance albeit slightly larger.
The slight difference ($\sim$10\%) between $d$ and $d^\mathrm{theo.}$ could be explained by higher density at the location of the second ring, different dust properties or uncertainties on the inclination and size estimations.
In this case we suggest that the outer ring is part of the outflow, see below.

\begin{table}
\caption{Geometrical determinations \modif{for the out-of-plane ring} \modif{of \object{HD101584} for both interpretations of Sect.\,\ref{sec:scenarios}.}}             
\label{tab:scenarios}      
\centering                          
\begin{tabular}{c c c c c }        
\hline\hline     \\     [-10pt]        
Interpretation & $h$ & $d$ & $d^\mathrm{theo.}$ & $\alpha_\mathrm{op}$  \\  
 & [au] & [au] & [au]  & [$^\circ$] \\[3pt] 
\hline \\     [-10pt]
A & 1.6 & 4.1 & 4.5 & 66  \\ [3pt]
B & 1.6 & 2.5 & 1.6 & 50  \\ [3pt]
\end{tabular}
\tablefoot{The distances are defined for a distance (D) of 1\,kpc.\\
$d^\mathrm{theo.}_\mathrm{cond.}$ is defined as $d^\mathrm{theo.}_\mathrm{cond.}$=d$_\mathrm{sub.}$ T$^2_\mathrm{sub.}$ / T$^2_\mathrm{cond.}$.}
\end{table}

\subsubsection{Interpretation B: Outer ring in the mid-plane}
\label{sec:ScB}

In this interpretation we assume that the outer ring lies in the mid-plane.
This would put the size of the disk inner rim to higher values, \modif{with a radius of} 3.7\,[D/1kpc]\,au, more compatible with most of the post-AGB circumbinary disk inner rims (Fig.\,\ref{fig:sketch}).
The height of the inner ring above the midplane is in this case 1.6\,[D/1kpc]\,au.
We then perform the same computations as for interpretation A but reversing the positions of the rings.
Therefore, the distance of the inner rim to the central star would be $d=\sqrt{(rD_1/2)^2 + h^2}$, which gives 2.5\,[D/1kpc]\,au.
As we did for interpretation A we can estimate the distance of the second ring to the central star assuming similar dust characteristics for both rings as $d^\mathrm{theo.} =  T_2^2 rD_2 / 2T_1^2  $.
For interpretation B, $d^\mathrm{theo.}_\mathrm{cond.} = 1.6$\,[D/1kpc]\,au, i.e., we find a larger difference ($\sim$56\%) \modif{compared to} the geometrical distance ($d$=2.5\,[D/1kpc]\,au) than for interpretation A.

\subsubsection{Interpretation A is more likely}

From our dataset we deduce that the inner ring represents 42\% of the total $H$-band flux and $\sim$58\% of the circumstellar flux.
Therefore it dominates the $H$-band emission. 
Moreover, a model of a disk with an inner radius at 2.15\,au successfully reproduces the SED at near-infrared wavelengths \citep{Olofsson2019}, which is very close to the inner ring radius we observe.
\modif{Further, the inner ring is more centered on the point source than the outer ring.}
These arguments strongly supports interpretation A.

In both interpretations we can estimate the geometrical distance from the star to the ring that is out of the mid-plane.
We also measure the temperatures of these rings.
Assuming an optically thin inner cavity and thermal equilibrium, the energy received by a ring at a certain distance of the star can be translated to a ring temperature using Eq.\,\ref{eq:Tsub}.
For interpretation A, the theoretical temperature is estimated to be 1068\,K which is remarkably close to the measured temperature ($T_2$=1014\,K).
For interpretation B the difference is larger as the theoretical temperature is 1743\,K, which is 200\,K above the measured one.
This result also makes interpretation A more likely.

\modif{Finally, the blue-shifted part of the large scale structure seen with ALMA is pointing in the west direction.
If the large and small-scale structures are linked, such an orientation agrees with interpretation A, where the out-of-plane structure is shifted in the west direction, and we do not detect its counter-part (in the red-shifted part) possibly because of the extinction caused by the disk.}

\subsection{Why do we not detect the companion?}
\label{sec:separation}

\begin{table*}
\caption{Parameters of the orbital separation taking into account the different scenarios.}             
\label{tab:binsep}      
\centering                          
\begin{tabular}{c c c c c c c c c c c}        
\hline\hline                 
Scenario &$P$ & $f(M)$ & $i$ & $a_1$ & $M_1$ & $M_2$ & $a_2$ & $a$ & d & $\rho$ \\ 
& [days] & [M$_\odot$] & [$^\circ$] & [au] & [M$_\odot$] & [M$_\odot$] & [au] & [au] & [kpc] & [mas]\\
\hline                        
post-RGB & 144 & 4$\times$10$^{-4}$ & 10 & 0.23 & 0.36 & 0.33 & 0.25 & 0.48 & 0.56 & 0.85 \\
post-RGB& 144 & 4$\times$10$^{-4}$ & 20 & 0.12 & 0.36 & 0.14 & 0.31 & 0.43 & 0.56 & 0.77 \\
post-AGB& 144 & 4$\times$10$^{-4}$ & 10 & 0.23 & 0.55 & 0.41 & 0.31 & 0.54 & 1.8 & 0.30 \\
post-AGB& 144 & 4$\times$10$^{-4}$ & 20 & 0.12 & 0.55 & 0.17 & 0.39 & 0.51 & 1.8 & 0.28 \\
\hline
post-RGB& 218 & 6$\times$10$^{-4}$ & 10 & 0.35 & 0.36 & 0.41 & 0.31 & 0.66 & 0.56 & 1.18 \\
post-RGB& 218 & 6$\times$10$^{-4}$ & 20 & 0.18 & 0.36 & 0.16 & 0.41 & 0.59 & 0.56 & 1.05 \\
post-AGB& 218 & 6$\times$10$^{-4}$ & 10 & 0.35 & 0.55 & 0.51 & 0.38 & 0.73 & 1.8 & 0.41 \\
post-AGB& 218 & 6$\times$10$^{-4}$ & 20 & 0.18 & 0.55 & 0.21 & 0.47 & 0.65 & 1.8 & 0.36 \\
\hline                                   
\end{tabular}
\end{table*}

\modif{Binarity is suggested by the periodicity seen in photometric and spectroscopic times series. \citet{Bakker1996} found a significant photometric period of 218$^d$ that is also tentatively found in radial velocities, but the low number of epochs and the line asymmetries could cause a bias in the obtained velocities. 
The fit was done for a circular orbit, but an eccentric one could have given better results.
Pulsation and stellar rotation were ruled out by \citet{Bakker1996} because of the period and amplitude of the radial velocity variations.
\citet{Bakker1996} argued that the photometric variability could be due to the periodic obscuration of the secondary continuum source (e.g., by a circum-secondary accretion disk). Given the orientation of the disk it would imply that the binary is strongly misaligned with the disk.
Another periodicity (144$^d$) was estimated by \citet{Diaz2007} from spectroscopic data, but the data were never shown. 
This period is not compatible with the one from \citet{Bakker1996}. 
It therefore remains possible that there is no secondary component in the system.
In the following we estimate the constrains we can bring from the non-detection of the companion assuming it is a binary system.
A low-mass stellar companion would not be detected due to the high contrast with the evolved star.
However, the near-infrared continuum emission at the position was detected in the case \modiff{of} the post-AGB binary system \object{IRAS08544-4431} \citep{Hillen2016} and was interpreted as coming from a circum-secondary accretion disk.
We therefore could also expect such a detection in \object{HD101584}.}

The evolutionary status of this object is unclear.
\citet{Olofsson2019} claim that the object is most likely a post-RGB binary rather than a post-AGB binary.
The different evolutionary scenarios lead to different distance and mass estimations and the two following scenarios were considered.
In the post-RGB case the system would be 0.56\,kpc \modif{away} with a present day primary mass ($M_1$) of 0.36\,M$_\odot$, whereas a post-AGB binary would be located at 1.8\,kpc and the primary would have a \modif{present day} mass of 0.55\,M$_\odot$.

The binary angular separation will be different in the two scenarios, and we want to investigate whether the non detection of the secondary can bring a constraint on its evolutionary status. 
From our observations, we found that the inclination of the disk is 20$^\circ$ whereas the analysis of the HGS seen in the ALMA data points towards an inclination of 10$^\circ$.
Hereafter we assume that the orbital plane \modif{of the putative companion} is in the disk plane and that the orbit is circular.
All those parameters influence the angular binary separation.
We have therefore computed the angular separations for each case.

We first estimated the mass function ($f(M)$) for each \modif{reported} period using:
\begin{equation}
    f(M) = \frac{K_1^3P}{2\pi G}
\end{equation}
where $K_1=3$\,km/s is the velocity semi-amplitude, $P$ the orbital period and $G$ the gravitational constant.
We then find the mass of the companion ($M_2$) by solving:
\begin{equation}
    f(M) = \frac{(M_2 \sin i)^3}{(M_1+M_2)^2}
\end{equation}
Then, we get the semimajor axis of the primary ($a_1$) via:
\begin{equation}
    a_1 = \sqrt[\leftroot{-1}\uproot{2}\scriptstyle 3]{\frac{f(M) G P^2}{4 \pi^2}} \frac{1}{\sin i}
\end{equation}
We compute the semimajor axis of the secondary ($a_2$) with:
\begin{equation}
    a_2 = \frac{M_1 a_1}{M_2}
\end{equation}
We obtain the physical separation by adding the two semimajor axes.
Finally, we use the distance ($D$) to get the angular separation ($\rho$):
\begin{equation}
    \rho = \frac{a}{D}
\end{equation}
The results for each case and each computation step are presented Table\,\ref{tab:binsep}.

The maximum angular distance is given by the post-RGB scenario with a period of 218$^d$ and an inclination of 10$^\circ$.
This separation of 1.2\,mas is of the same order as our maximum resolution power: $\lambda/2B_\mathrm{max} \sim 1.3$\,mas.
Our observations are therefore not constraining the binary separation and therefore can not exclude any of the two evolutionary scenarios proposed in \citet{Olofsson2019}, irrespective of its contribution to the $H$-band flux.

\section{Conclusions}
\label{sec:ccl}

The conclusions we have obtained from the \object{HD101584} VLTI/PIONIER data analysis are:

   \begin{enumerate}
      \item \modif{The reconstructed image from} VLTI/PIONIER observations of the environment \modif{of \object{HD101584}} \modif{reveals} an unexpected double ring morphology.
      \item Both rings are compatible with a dust sublimation rim but at different temperatures.
      \item The two rings are not concentric. \modif{We postulate that t}he shift between the two rings is compatible with a projection effect where one of the two rings is not in the mid-plane. The height ($h$) of the ring which is out of the mid-plane is 1.6[D/1kpc]\,au. The ring which is out of the plane may be linked with the hourglass shaped outflow seen at large scale with ALMA.
      \item The origin of the out-of-plane ring is unclear. It is shifted in the direction of the blue-shifted part of the outflow seen by ALMA, i.e., the part of the outflow moving towards us. We speculate that it is tracing either a density enhancement or the dust condensation front in the outflow. We propose additional infrared interferometric observations of this target to discriminate between these two hypotheses.
      \item The non-detection of the secondary is compatible with post-RGB\modif{,} post-AGB \modif{and merger} evolutionary scenarios of this object.
   \end{enumerate}

High-angular observations in the infrared are able to image the stellar environments at an astronomical\modif{-}unit scale allowing us to probe important processes in the evolution of binary stars \modif{or mergers}.
The imaging approach is crucial to reveal unexpected features and guide the interpretation and modeling of a complex \modif{circumstellar and} circumbinary environment.
The secondary ring detected in \object{\object{HD101584}} is certainly unexpected. 
Thanks to observations at larger scales, one can now study these objects thoroughly and link morphological features observed at different scales (HST, ALMA) to constrain the complex physical processes at place in circum\modif{stellar} environments.
To \modif{confirm our out-of-plane scenario,} we propose to re-observe this target at the same wavelengths but also at longer infrared wavelengths (with GRAVITY or MATISSE at the VLTI) to probe its density structure and understand its launching mechanism and impact on the fate of the system.

\begin{acknowledgements}
	We want to thank the referee for comments that improved the manuscript.
     JK and HVW acknowledge support from the research council of the KU Leuven under grant number C14/17/082.
     HO and WV acknowledge financial support from the Swedish Research Council.
\end{acknowledgements}

\bibliographystyle{aa} 
\bibliography{references} 

\begin{appendix}

\section{Log of the observations}
\label{app:log}

The log of the observations is indicated Table\,\ref{tab:log}.

\begin{table}
\caption{Log of observations}             
\label{tab:log}      
\centering                          
\begin{tabular}{c c c c}        
\hline\hline                 
Run & Date & MJD & Configuration \\    
\hline                        
099.D-0088(A) & 2017-04-10 & 57853.2 & A0-B2-C1-D0\\
099.D-0088(A) & 2017-04-10 & 57853.2 & A0-B2-C1-D0\\
099.D-0088(A) & 2017-04-10 & 57854.0 & A0-B2-C1-D0\\
099.D-0088(A) & 2017-04-11 & 57854.0 & A0-B2-C1-D0\\
099.D-0088(A) & 2017-04-11 & 57854.0 & A0-B2-C1-D0\\
099.D-0088(A) & 2017-04-11 & 57854.0 & A0-B2-C1-D0\\
099.D-0088(A) & 2017-04-11 & 57854.1 & A0-B2-C1-D0\\
099.D-0088(A) & 2017-04-11 & 57854.1 & A0-B2-C1-D0\\
099.D-0088(A) & 2017-04-11 & 57854.1 & A0-B2-C1-D0\\
099.D-0088(A) & 2017-04-11 & 57854.1 & A0-B2-C1-D0\\
099.D-0088(A) & 2017-04-11 & 57854.1 & A0-B2-C1-D0\\
099.D-0088(A) & 2017-04-11 & 57854.1 & A0-B2-C1-D0\\
099.D-0088(A) & 2017-04-11 & 57854.2 & A0-B2-C1-D0\\
099.D-0088(A) & 2017-04-11 & 57854.2 & A0-B2-C1-D0\\
099.D-0088(A) & 2017-04-11 & 57854.2 & A0-B2-C1-D0\\
099.D-0088(A) & 2017-04-11 & 57854.2 & A0-B2-C1-D0\\
099.D-0088(C) & 2017-04-22 & 57865.1 & D0-G2-J3-K0\\
099.D-0088(C) & 2017-04-22 & 57865.1 & D0-G2-J3-K0\\
099.D-0088(C) & 2017-04-22 & 57865.1 & D0-G2-J3-K0\\
099.D-0088(C) & 2017-04-22 & 57865.2 & D0-G2-J3-K0\\
099.D-0088(C) & 2017-04-22 & 57865.2 & D0-G2-J3-K0\\
099.D-0088(C) & 2017-04-22 & 57865.2 & D0-G2-J3-K0\\
099.D-0088(C) & 2017-04-24 & 57867.1 & D0-G2-J3-K0\\
099.D-0088(C) & 2017-04-24 & 57867.1 & D0-G2-J3-K0\\
099.D-0088(C) & 2017-04-24 & 57867.2 & D0-G2-J3-K0\\
099.D-0088(C) & 2017-04-24 & 57867.2 & D0-G2-J3-K0\\
099.D-0088(C) & 2017-04-25 & 57868.1 & A0-G2-D0-J3\\
099.D-0088(C) & 2017-04-25 & 57868.1 & A0-G2-D0-J3\\
099.D-0088(B) & 2017-04-26 & 57869.2 & A0-G1-J2-J3\\
099.D-0088(B) & 2017-04-26 & 57869.2 & A0-G1-J2-J3\\
099.D-0088(B) & 2017-04-26 & 57869.2 & A0-G1-J2-J3\\
099.D-0088(B) & 2017-04-27 & 57870.1 & A0-G1-J2-J3\\
099.D-0088(B) & 2017-04-27 & 57870.2 & A0-G1-J2-J3\\
099.D-0088(B) & 2017-04-28 & 57871.1 & A0-G1-J2-J3\\
099.D-0088(B) & 2017-04-28 & 57871.1 & A0-G1-J2-J3\\
099.D-0088(B) & 2017-04-29 & 57872.1 & A0-G1-J2-J3\\
099.D-0088(B) & 2017-04-29 & 57872.1 & A0-G1-J2-J3\\
099.D-0088(B) & 2017-04-29 & 57872.1 & A0-G1-J2-J3\\
099.D-0088(B) & 2017-04-29 & 57872.1 & A0-G1-J2-J3\\
099.D-0088(B) & 2017-04-29 & 57872.1 & A0-G1-J2-J3\\
099.D-0088(B) & 2017-04-29 & 57872.2 & A0-G1-J2-J3\\
099.D-0088(B) & 2017-04-29 & 57872.2 & A0-G1-J2-J3\\
099.D-0088(B) & 2017-04-29 & 57872.2 & A0-G1-J2-J3\\
099.D-0088(B) & 2017-04-30 & 57873.0 & A0-G1-J2-J3\\ 
099.D-0088(B) & 2017-04-29 & 57873.0 & A0-G1-J2-J3\\
099.D-0088(B) & 2017-04-29 & 57873.0 & A0-G1-J2-J3\\
099.D-0088(B) & 2017-04-30 & 57873.0 & A0-G1-J2-J3\\
099.D-0088(C) & 2017-05-19 & 57892.1 & G2-J3-K0\\
099.D-0088(C) & 2017-05-19 & 57892.1 & G2-J3-K0\\
099.D-0088(C) & 2017-05-19 & 57892.1 & G2-J3-K0\\
099.D-0088(C) & 2017-05-20 & 57893.1 & D0-G2-J3-K0\\
099.D-0088(C) & 2017-05-20 & 57893.1 & D0-G2-J3-K0\\
\hline                                   
\end{tabular}
\end{table}

\section{Chromatic parameters determination}
\label{app:chrom}

To determine the optimal chromatic parameters we made a grid of 50$\times$50 pairs of \fso and \denv.
For each pair of parameters we reconstructed an image, and we selected the pair for which the cost $f$ is the lowest.
We varied \fso between 0.15 and 0.35 and \denv between -1 and 5.

The second image reconstruction presented in Sect.\,\ref{sec:Modelfit} was done setting the stellar-to-total flux ratio (\fso) to 0.28 and making a grid on the spectral index of the environment (\denv). 
This linear grid was made of 200 values between -3.1 and 4.9 (corresponding to black body temperatures between the one of the star and 880\,K).
We find \denv=1.75 as the most likely value which is compatible with the grid on both chromatic parameters made in the first step (Fig.\,\ref{fig:chrom}). 

\section{Effect of the regularization weight}
\label{sec:mu}

 \begin{figure}
 \centering
  \includegraphics[width=9cm]{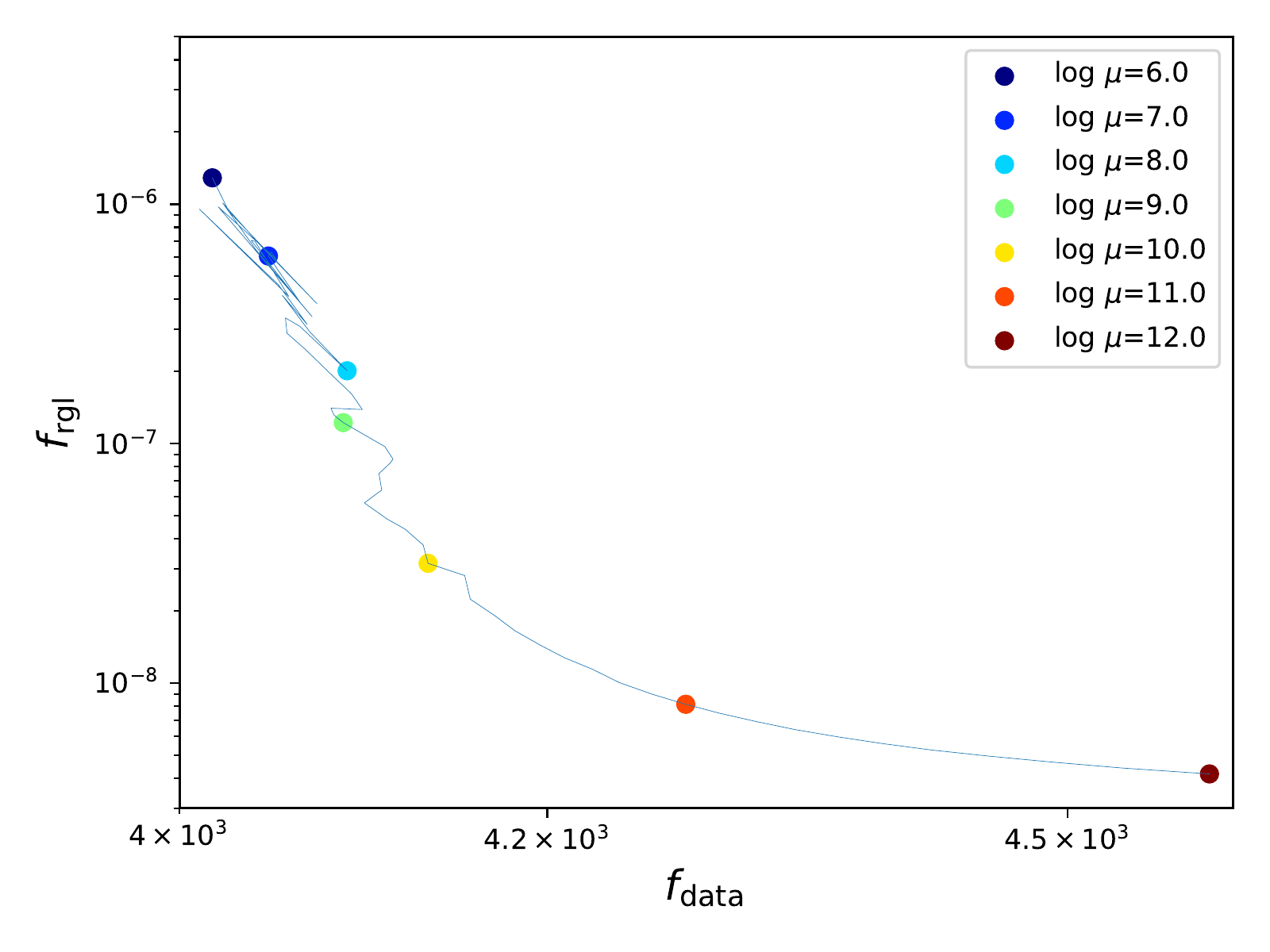}
 \caption{The L-curve for the image reconstruction displayed in Fig.\,\ref{fig:image}.}
 \label{fig:Lcu}%
 \end{figure}

The L-curve used in the image reconstruction of Sect.\,\ref{sec:ImgRec} is on Fig.\,\ref{fig:Lcu}.
We distinguish the two regimes of the L curve.
In the low $\mu$ case the L curve is not smooth probably because of local minimas due to the interferometric data (e.g. not convexity due to phases).

 \begin{figure*}
 \centering
  \includegraphics[width=19cm]{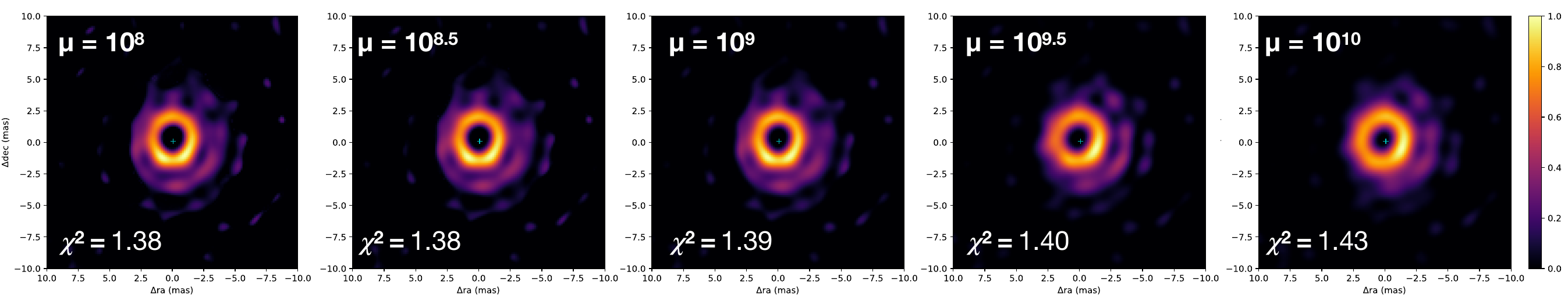}
 \caption{Image reconstructions using different regularization weights. The images are made using \fso=0.28 and \denv=1.75.}
 \label{fig:ImageVsMu}%
 \end{figure*}

To illustrate our exploration of the regularization weight parameter we show Fig.\,\ref{fig:ImageVsMu} how the image changes with the regularization weight for a constant \fso (\fso=0.28).
As an illustration we also indicated the reduced $\chi^2$ (even thought it is not the value that is reduced when optimizing the image) that increases for higher regularization weights ($\mu > 10^9$) as expected when switching from a regime that minimizes the likelihood to a regime dominated by the regularization cost.

We can see that the global characteristics of the image are retrieved, namely a bright inner ring and an extended structure looking like an outer ring slightly shifted towards the West.
For higher regularization weights the outer ring becomes an arc in the West direction.
Because of the smoothness regularization used the emission from the Eastern side of the ring became super-imposed with the inner rim.
Images with higher $\mu$ do not change our interpretation as we can still model the data with the double ring morphology.

 \begin{figure*}
 \centering
  \includegraphics[width=19cm]{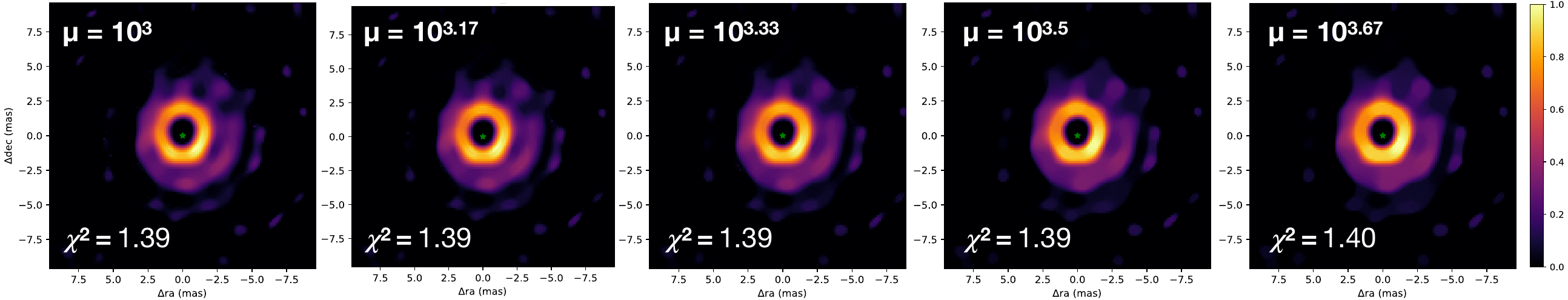}
 \caption{\modif{Image reconstructions using the total variation regularization with different regularization weights. The images are made using \fso=0.28 and \denv=1.75.}}
 \label{fig:ImageVsMuTV}%
 \end{figure*}
\modif{We have also tested the effect of a different regularization. We used the total variation regularization \citep{Renard2010}. This regularization reduce\modif{s} the number of gradients in the image creating plateaus in the image. This regularization is not quadratic and is more likely to fall in a local minimum than the quadratic smoothness regularization. The images obtained with this regularization show similar features (see Fig.\,\ref{fig:ImageVsMuTV}), namely the inner ring and the secondary ring that is closer to the inner ring on the East side.}

\begin{figure*}
 \centering
  \includegraphics[width=6cm]{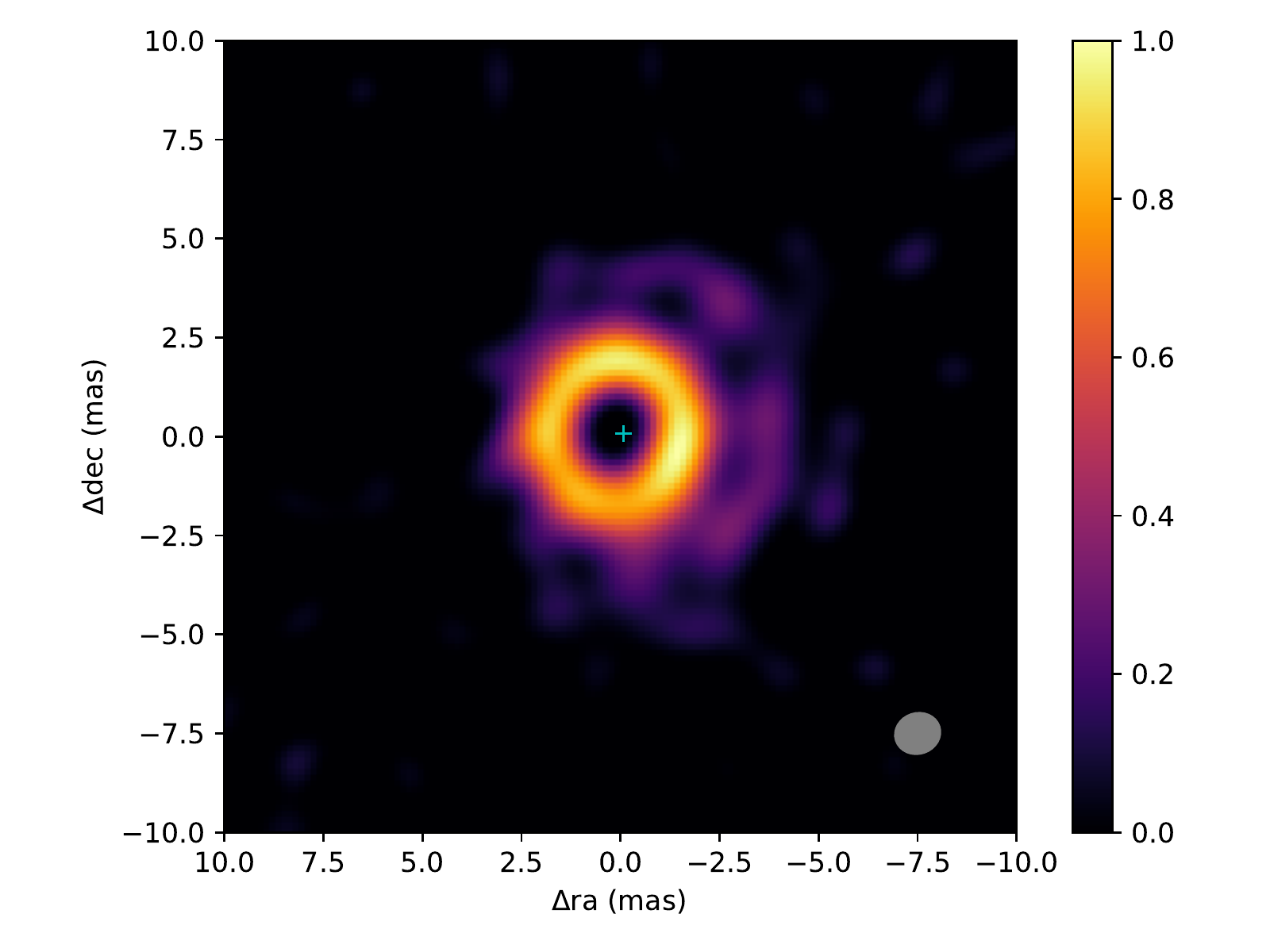}
  \includegraphics[width=6cm]{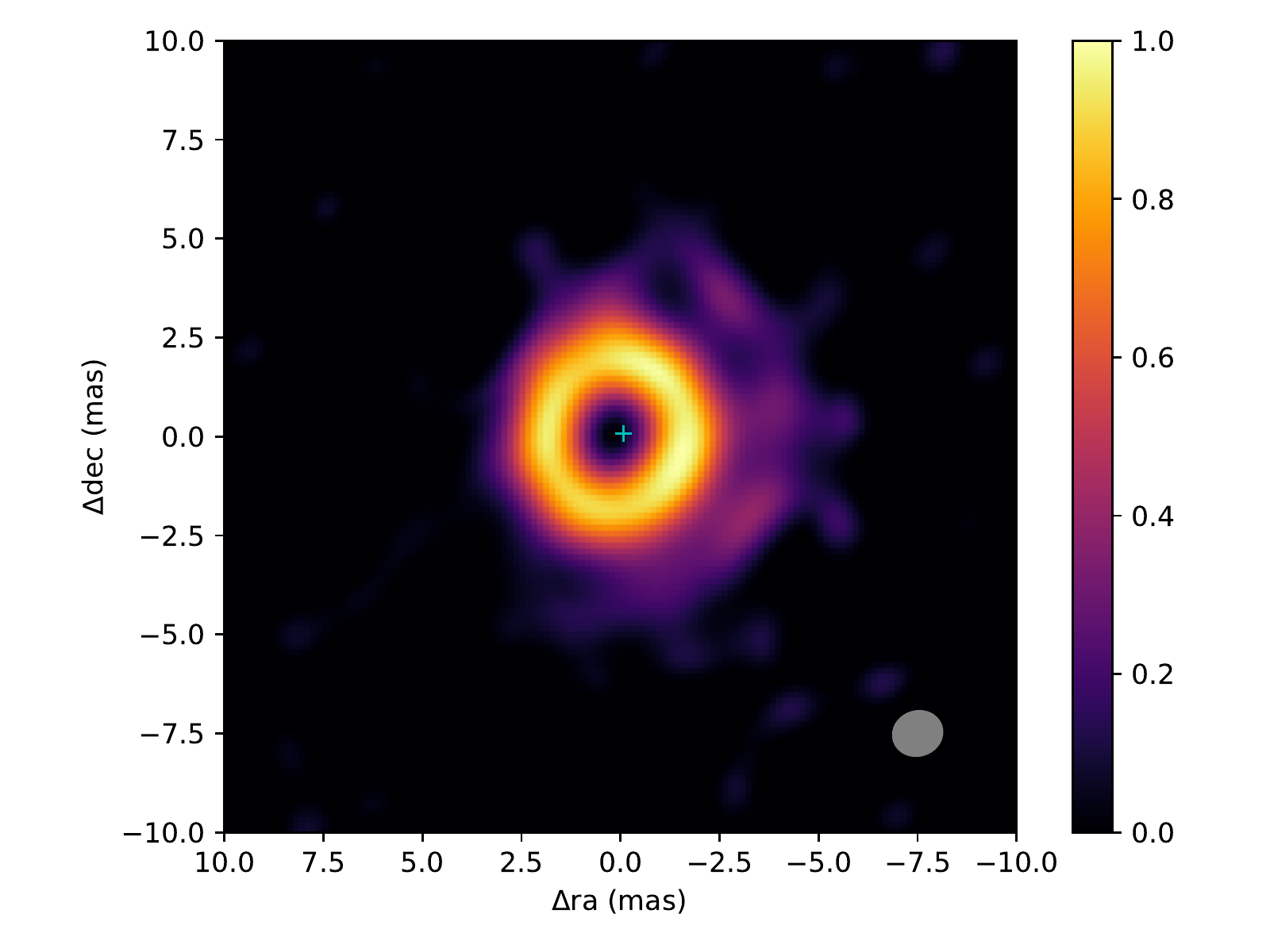}
 \caption{\modif{Image reconstructions using the quadratic smoothness regularization with the three shortest-wavelength channels (left) and the three longest-wavelength ones (right). The images are made using \fso=0.28 and \denv=1.75.}}
 \label{fig:Imageshortlong}%
 \end{figure*}

\modif{Finally, Fig.\,\ref{fig:Imageshortlong} shows image reconstructions when splitting the dataset between the channels with shorter wavelengths ($\lambda<$1.65$\mu$m) and those with longer ones ($\lambda>$1.65$\mu$m). More artifacts are expected because we divide the number of data points by a factor of two and the beam size is different between the images (the image with shorter wavelengths will have a better angular resolution). We recover the ring in both images together with the west-side arc, although beam size effects are visible. The east-side part of the outer ring is seen for the shortest wavelengths only, where higher angular resolution is needed, even though its structure is distorted. }

\section{Fit of the images to the dataset}

\begin{figure}
 \centering
  \includegraphics[width=10cm]{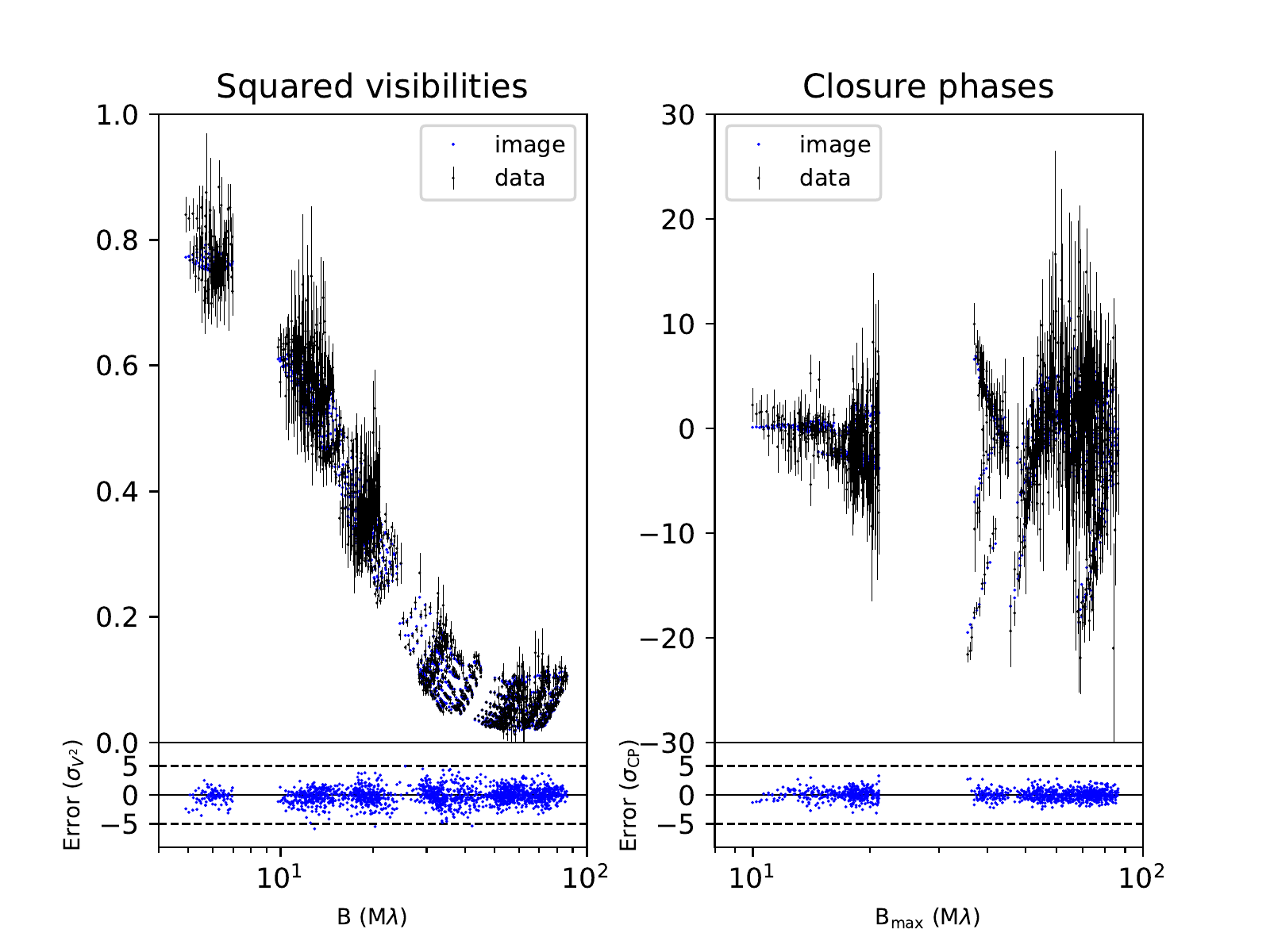}
 \caption{Comparison of the observables from the image reconstruction of \object{HD101584} done in Sect.\,\ref{sec:ImgRec} with chromatic parameters being \fso=0.244 and \denv=2.45. Left: The V2 of the data (black) and from the image reconstruction (blue). Right: The CP of the data (black) and the image reconstruction (blue). Top: The absolute values. Bottom: the residuals. For clarity the x-axis in log-scale.}
 \label{fig:Image1vsdata}%
 \end{figure}

\begin{figure}
 \centering
  \includegraphics[width=10cm]{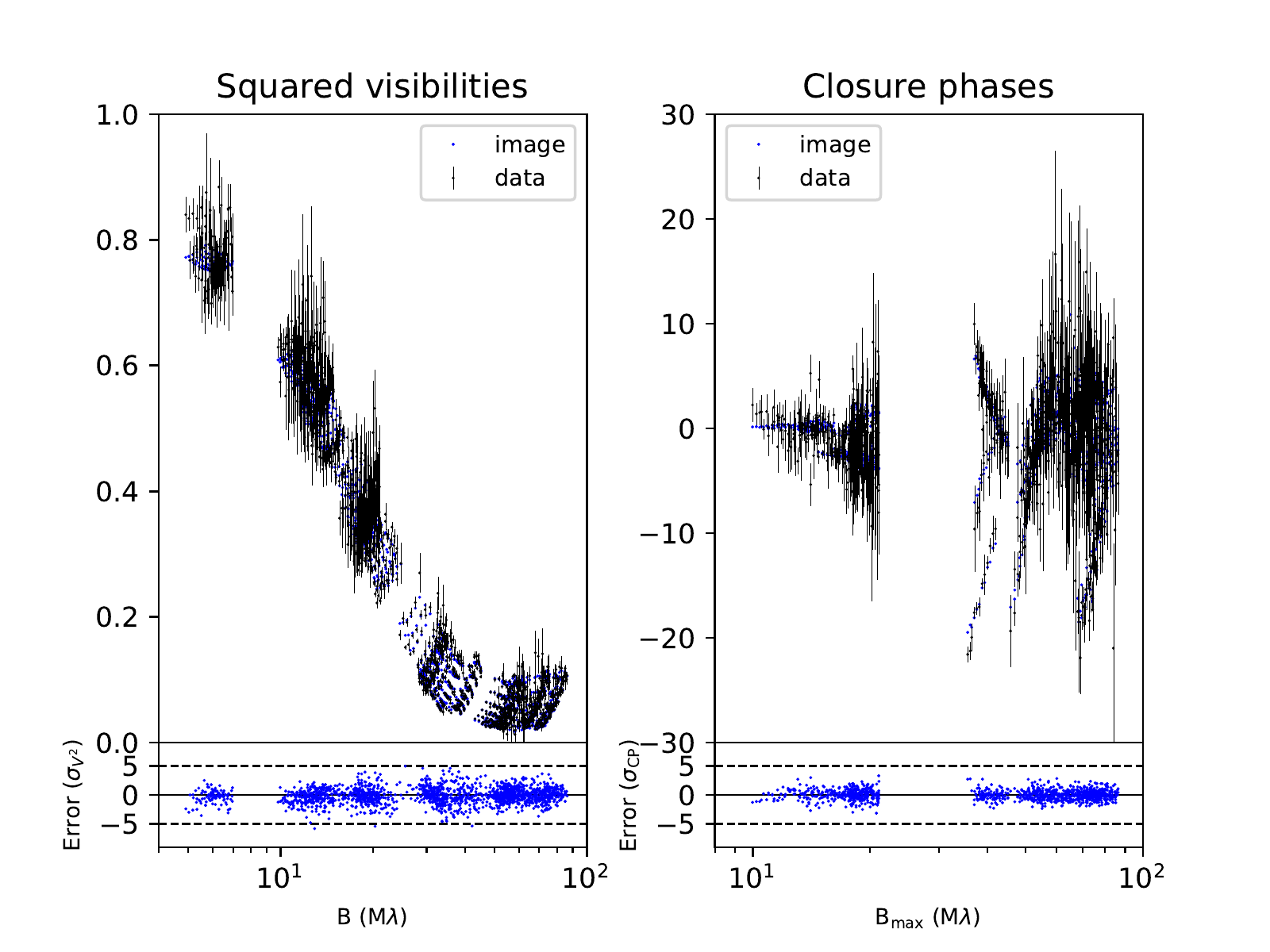}
 \caption{Comparison of the observables from the image reconstruction of \object{HD101584} done in Sect.\,\ref{sec:Modelfit} with chromatic parameters being \fso=0.28 and \denv=1.75. Left: The V2 of the data (black) and from the image reconstruction (blue). Right: The CP of the data (black) and the image reconstruction (blue). Top: The absolute values. Bottom: the residuals. For clarity the x-axis in log-scale.}
 \label{fig:Image2vsdata2}%
 \end{figure}

The comparison of the interferometric observables from the images to the dataset are shown in Fig.\,\ref{fig:Image1vsdata} and \ref{fig:Image2vsdata2} for the image reconstruction with \fso=0.244 and \denv=2.45, and \fso=0.28 and \denv=1.75 respectively.

\section{Geometrical model}
\label{app:model}

The geometrical model is inspired by the reconstructed image.
Because of the linearity of the Fourier transform the model can be defined as the flux weighted sum of individual components.
It is made of a point source to reproduce the star, two rings and a background to take into account the over-resolved flux.
The two rings orientation is defined with the same inclination and position angle.
Moreover, both the point source and the outer ring can be shifted w.r.t. the inner ring.
Hereafter, we mathematically describe the geometrical model we used to reproduce the interferometric data.

To combine the different components of the model one need to defined the fluxes of each component. 
Those fluxes are normalized and positive. They have to respect the following constraints:
\begin{eqnarray}
\sum_\mathrm{i} f^\mathrm{i}_0 &=& 1 \label{eqn:norm}\\
f^\mathrm{i}_0 &\geqslant& 0 \\
f^\mathrm{i}_0 &\leqslant& 1 ,
\end{eqnarray}
where $f^\mathrm{i}_0$ is the flux ratio of the $i$-th component (here it can be the star, one of the rings or the background flux) at 1.65$\mu$m.
To extrapolate the flux ratios to other wavelengths ($f^\mathrm{i}$) probed by the data, spectral laws are assigned to each components.

For both rings a black body at temperature ($T_\mathrm{i}$) is assigned such as:
\begin{equation}
f^\mathrm{i} = f^\mathrm{i}_0 \Bigg(\frac{BB(\lambda,T_\mathrm{i})}{BB(1.65\mu\mathrm{m},T_\mathrm{i})}\Bigg),
\end{equation}
where $BB$ is the Planck black body function and $\lambda$ is the wavelength of an observation.

For the point source and the back-ground the spectrum is a power-law with a spectral index ($d_\mathrm{i} = \frac{d \log F_\lambda}{d \log \lambda}$) such as:
\begin{equation}
f^\mathrm{i} = f^\mathrm{i}_0 \Bigg(\frac{\lambda}{1.65\mu\mathrm{m}}\Bigg)^\mathrm{d_\mathrm{i}},
\end{equation}

In the following we describe the analytical Fourier transforms we used in the model.
We describe each component separately.
\begin{itemize}
\item \textbf{The star:}
The visibility of a point source equals unity for every spatial frequency ($V^*=1$). The star can be shifted w.r.t. the inner ring center by $\Delta x$ and $\Delta y$ for right ascension and declination shift respectively. 
Therefore, the complex visibility of the star is:
\begin{equation}
 V^*(u,v) = \exp-2 i \pi (u \Delta x + v \Delta y),
\label{eqn:Vstar}
\end{equation}
where $u$ and $v$ are the spatial frequencies in the West-East and South-North directions.
\item \textbf{The ring:}
The ring is first defined as an infinitesimal ring distribution. Its visibility ($V^\mathrm{ring0}(u,v)$) equals to:
\begin{equation}
    V^\mathrm{ring0}(u,v) = J_\mathrm{0}(\pi\rho'rD_\mathrm{i}),
\end{equation}
where $J_\mathrm{0}$ is the Bessel function of the 0$^\mathrm{th}$ order, $rD_\mathrm{i}$ the diameter of the $i$-th ring and $\rho'$ is the spatial frequency of a data point corrected for inclination ($i$) and position angle ($PA$) of the ring such as:
\begin{eqnarray}
    \rho' &=& \sqrt{u'^2 + v'^2}\\
    u' &=& u \cos PA + v \sin PA \\
    v' &=& (-u \sin PA + v \cos PA) \cos i.
\end{eqnarray}
The ring is then convolved by a Gaussian such as:
\begin{equation}
    V^\mathrm{ring}(u,v) = V^\mathrm{ring0}(u,v) \exp \frac{-\bigg(\pi \frac{rD_\mathrm{i}}{2} rW_\mathrm{i} \sqrt{u^2 + v^2} \bigg)^2 }{4 \ln 2}, \label{eqn:ring}
\end{equation}
where $rW_\mathrm{i}$ is the ratio of the $i$-th ring full width at half maximum to its radius.

Additionally, for the second ring a phasor is added for its shift:
\begin{equation}
    V^\mathrm{ring2}(u,v) = V^\mathrm{ring}(u,v) \exp (-2i \pi (\Delta x_\mathrm{ring2} u + \Delta y_\mathrm{ring2} v ) ) , \label{eqn:ring2}
\end{equation}
with the ring shifts $\Delta x_\mathrm{ring2}$ and $\Delta y_\mathrm{ring2}$ are defined as:
\begin{eqnarray}
    \Delta x_\mathrm{ring2} &=& \Delta R_\mathrm{ring2} \cos PA \\
    \Delta y_\mathrm{ring2} &=& \Delta R_\mathrm{ring2} \sin PA
     \label{eqn:ring2shift}
\end{eqnarray}
with $\Delta R_\mathrm{ring2}$ being the ring2 shift.
Ring2 can therefore only be shifted in the direction of the minor axis.

\item \textbf{The background:}
The extended flux is modelled by an over-resolved emission that has a null visibility. 

\item \textbf{The final model:}
As the flux ratios of the three components are normalized to 1 at 1.65$\mu$m (Eqn.\ref{eqn:norm}), $f^\mathrm{ring1}_0$ is defined as:
\begin{equation}
    f^\mathrm{ring1}_0 = 1 - f^\mathrm{*}_0 - f^\mathrm{ring2}_0  - f^\mathrm{bg}_0.
\end{equation}
The final visibility can therefore be written as a linear combination of the four components of the model such as:
\begin{equation}
    V^\mathrm{tot}(u,v) = \frac{f^\mathrm{*} V^*(u,v) + f^\mathrm{ring1} V^\mathrm{ring1}(u,v) + f^\mathrm{ring2} V^\mathrm{ring2}(u,v) }{ f^\mathrm{*} + f^\mathrm{ring1} + f^\mathrm{ring2} + f^\mathrm{bg} }.
\end{equation}

\end{itemize}

\begin{figure*}
    \centering
    \includegraphics[width=17cm]{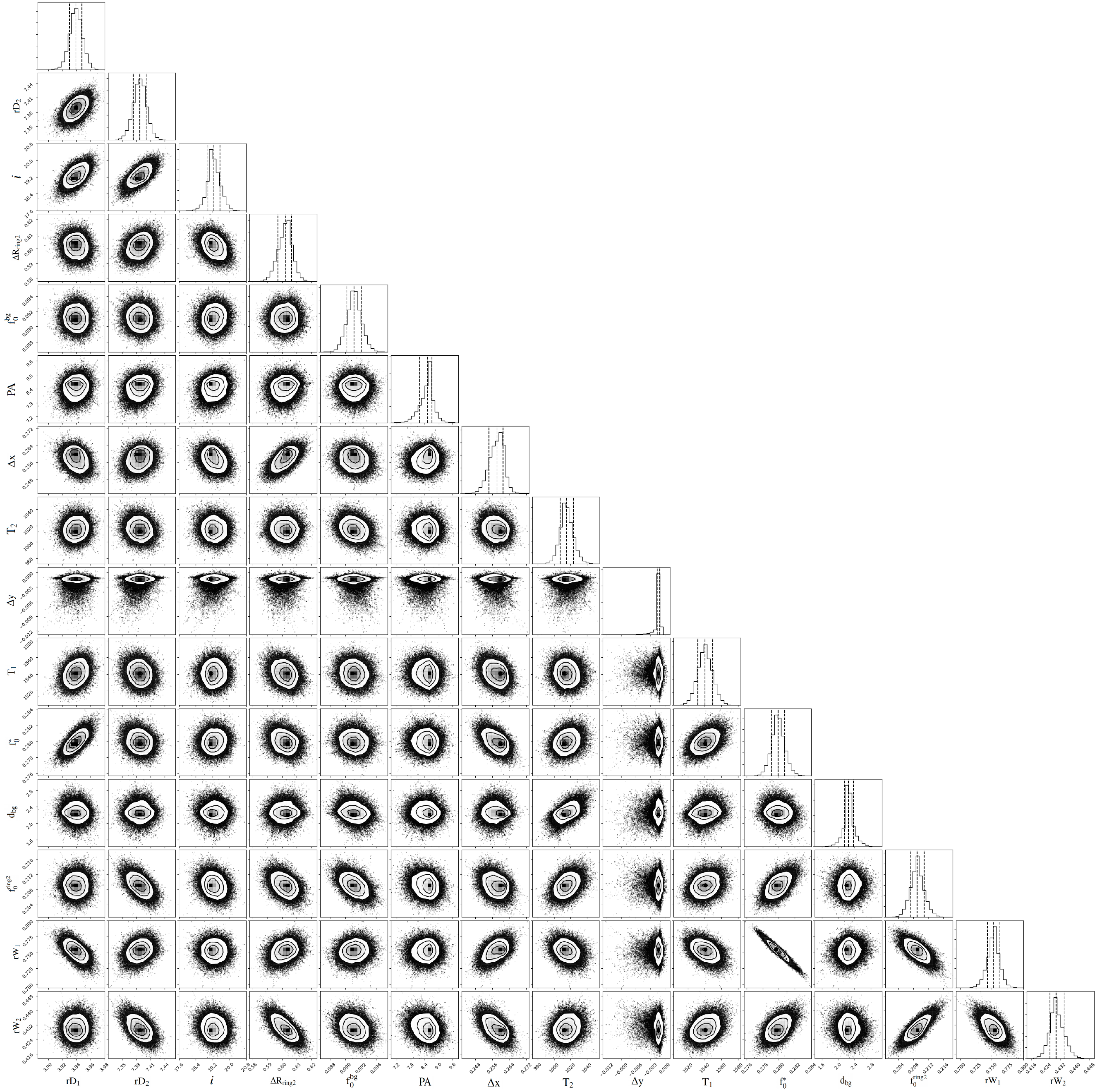}
    \caption{\modiff{Corner plot from the MCMC fit.}}
    \label{fig:MCMC}
\end{figure*}

\begin{figure}
 \centering
  \includegraphics[width=9cm]{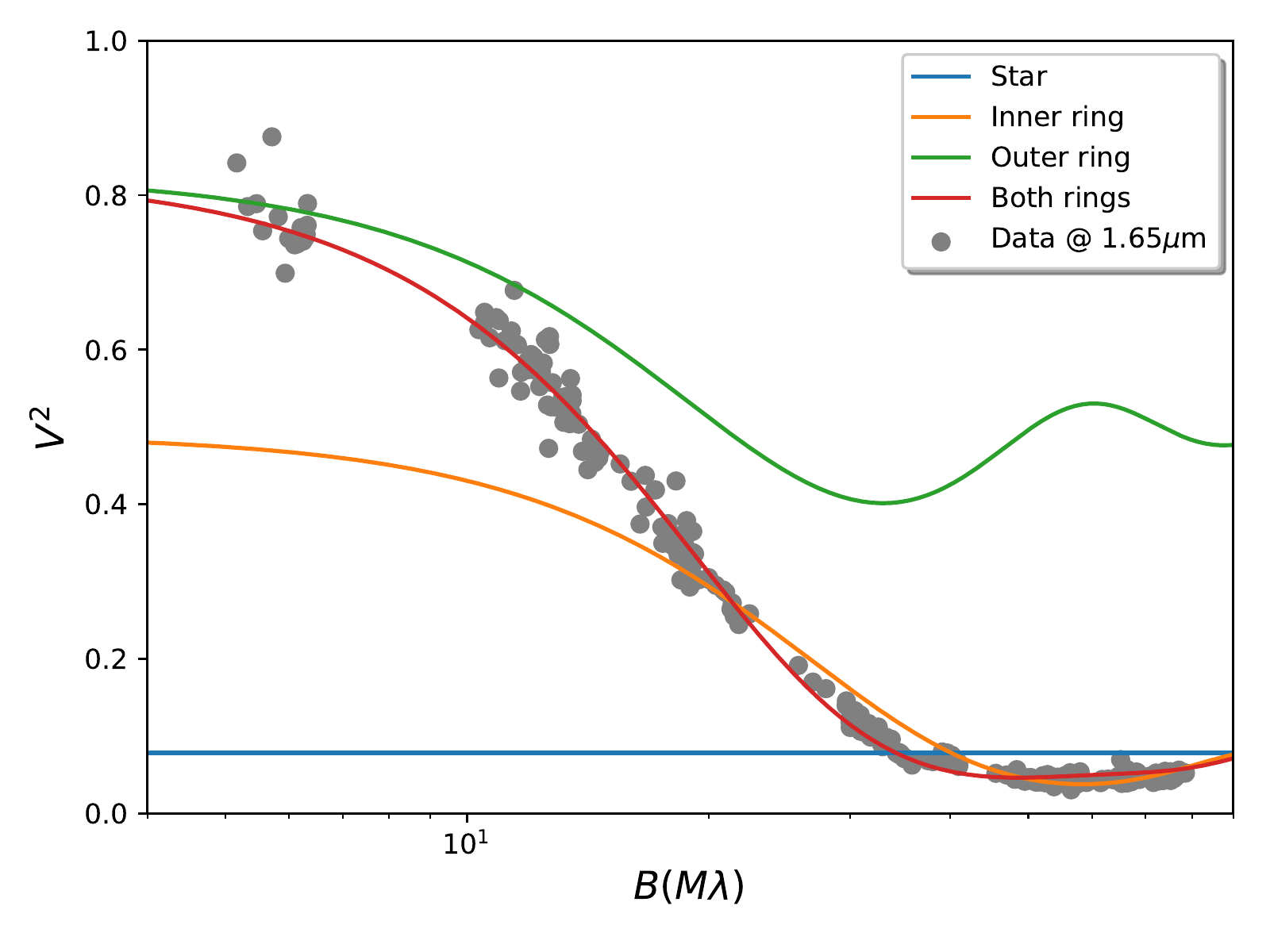}
 \caption{\modif{Plotting the Bessel functions of the rings and the star from the best-fit geometric model over data at 1.65$\mu$m. Second-order variations due to shift of the ring are not taken into account. The outer ring is plotted with a shift corresponding to the fluxes of the inner ring and the star flux ratios. The inner ring is plotted with a shift corresponding to the stellar flux ratio. There is no strong features in the visibilities at the location of the gaps in spatial frequencies. The spatial frequencies are in log scale.}}
 \label{fig:bessel}%
 \end{figure}

\end{appendix}

\end{document}